%% file: paperInfn.tex
\documentclass[10pt,a4paper,titlepage,twocolumn,dvips,showpacs,preprint,superscriptaddress,nofootinbib]{revtex4}
\usepackage[hmargin=2cm,vmargin={2cm,1.5cm},includefoot]{geometry}
\usepackage{color}
\usepackage{graphicx}
\usepackage[T1]{fontenc}
\usepackage[latin1]{inputenc}
\usepackage{txfonts}
\usepackage{relsize}
\usepackage{enumerate}
\usepackage{array}
\usepackage{tabularx}
\usepackage{amstext}

\input babarsym

\graphicspath{{figure/}}

\makeatletter
\DeclareRobustCommand\bfseries{%
  \not@math@alphabet\bfseries\mathbf
  \fontseries\bfdefault\selectfont\boldmath}

\makeatother

\newcommand\SuperB{\textsl{SuperB}\xspace}
\newcommand\Belle{\textit{Belle}\xspace}

\newcommand{\infnHeader}{
  \begin{tabular}{@{}cc@{}}
    \begin{tabular}{@{}c@{}}
      \includegraphics[width=80pt]{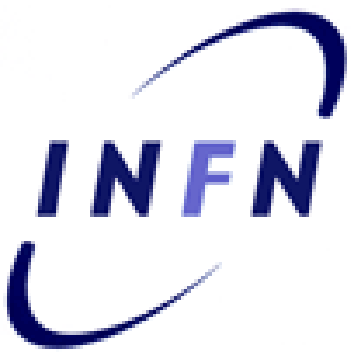}
    \end{tabular}
    &
    \begin{tabular}{c}
      \LARGE\sffamily ISTITUTO NAZIONALE DI FISICA NUCLEARE \\
    \end{tabular}
  \end{tabular}
  \vspace{0.5cm}
  \rule{16.0cm}{0.09mm} \\
  \begin{flushright}
    {\underline{\bf INFN-AE 05-08}} \\
    {\small\bf 20 Dicembre 2005} \\
  \end{flushright}
}

\begin{document}

\makeatletter
\def\@pacs@name{PACS: }%
\makeatother
  

\title{%
  \infnHeader
  {\larger\larger
    {\smaller\smaller INFN Roadmap Report} \\[1ex]
    {\larger \SuperB: a linear high-luminosity $B$ Factory}}
}

\iftrue

\author{J.~Albert}
\affiliation{California Institute of Technology, Pasadena, CA 91125, USA }
\author{S.~Bettarini}
\affiliation{Universit\`a di Pisa, Dipartimento di Fisica, Scuola Normale Superiore and INFN, I-56127 Pisa, Italy }
\author{M.~Biagini}
\affiliation{Laboratori Nazionali di Frascati dell'INFN, I-00044 Frascati, Italy }
\author{G.~Bonneaud}
\affiliation{Ecole Polytechnique, LLR, F-91128 Palaiseau, France }
\author{Y.~Cai}
\affiliation{Stanford Linear Accelerator Center, Stanford, CA 94309, USA }
\author{G.~Calderini}
\affiliation{Universit\`a di Pisa, Dipartimento di Fisica, Scuola Normale Superiore and INFN, I-56127 Pisa, Italy }
\author{M.~Ciuchini}
\affiliation{Universit\`a di Roma Tre, Dipartimento di Fisica and INFN, I-00146 Roma, Italy }
\author{G.~P.Dubois-Felsmann}
\affiliation{California Institute of Technology, Pasadena, CA 91125, USA }
\author{S.~Ecklund}
\affiliation{Stanford Linear Accelerator Center, Stanford, CA 94309, USA }
\author{F.~Forti}
\affiliation{Universit\`a di Pisa, Dipartimento di Fisica, Scuola Normale Superiore and INFN, I-56127 Pisa, Italy }
\author{T.~J.~Gershon}
\affiliation{Department of Physics, University of Warwick, Coventry CV4 7AL, United Kingdom }
\author{M.~A.~Giorgi}
\affiliation{Universit\`a di Pisa, Dipartimento di Fisica, Scuola Normale Superiore and INFN, I-56127 Pisa, Italy }
\author{D.~G.~Hitlin}
\affiliation{California Institute of Technology, Pasadena, CA 91125, USA }
\author{D.~W.~G.~S.~Leith}
\affiliation{Stanford Linear Accelerator Center, Stanford, CA 94309, USA }
\author{A.~Lusiani}
\affiliation{Universit\`a di Pisa, Dipartimento di Fisica, Scuola Normale Superiore and INFN, I-56127 Pisa, Italy }
\author{D.~B.~MacFarlane}
\affiliation{Stanford Linear Accelerator Center, Stanford, CA 94309, USA }
\author{F.~Martinez-Vidal}
\affiliation{IFIC, Universitat de Valencia-CSIC, E-46071 Valencia, Spain }
\author{N.~Neri}
\affiliation{Universit\`a di Pisa, Dipartimento di Fisica, Scuola Normale Superiore and INFN, I-56127 Pisa, Italy }
\author{A.~Novokhatski}
\affiliation{Stanford Linear Accelerator Center, Stanford, CA 94309, USA }
\author{M.~Pierini}
\affiliation{University of Wisconsin, Madison, Wisconsin 53706, USA }
\author{G.~Piredda}
\affiliation{Universit\`a di Roma La Sapienza, Dipartimento di Fisica and INFN, I-00185 Roma, Italy }
\author{S.~Playfer}
\affiliation{University of Edinburgh, Edinburgh EH9 3JZ, United Kingdom }
\author{F.~C.~Porter}
\affiliation{California Institute of Technology, Pasadena, CA 91125, USA }
\author{P.~Raimondi}
\affiliation{Laboratori Nazionali di Frascati dell'INFN, I-00044 Frascati, Italy }
\author{B.~N.Ratcliff}
\affiliation{Stanford Linear Accelerator Center, Stanford, CA 94309, USA }
\author{A.~Roodman}
\affiliation{Stanford Linear Accelerator Center, Stanford, CA 94309, USA }
\author{J.~Seeman}
\affiliation{Stanford Linear Accelerator Center, Stanford, CA 94309, USA }
\author{L.~Silvestrini}
\affiliation{Universit\`a di Roma La Sapienza, Dipartimento di Fisica and INFN, I-00185 Roma, Italy }
\author{A.~Stocchi}
\affiliation{Laboratoire de l'Acc\'el\'erateur Lin\'eaire, F-91898 Orsay, France }
\author{M.~Sullivan}
\affiliation{Stanford Linear Accelerator Center, Stanford, CA 94309, USA }
\author{U.~Wienands}
\affiliation{Stanford Linear Accelerator Center, Stanford, CA 94309, USA }
\author{W.~J.~Wisniewski}
\affiliation{Stanford Linear Accelerator Center, Stanford, CA 94309, USA }

\else 

\author{J.~Albert}
\author{G.~P.Dubois-Felsmann}
\author{D.~G.~Hitlin}
\author{F.~C.~Porter}
\affiliation{California Institute of Technology, Pasadena, CA 91125, USA }
\author{G.~Bonneaud}
\affiliation{Ecole Polytechnique, LLR, F-91128 Palaiseau, France }
\author{S.~Playfer}
\affiliation{University of Edinburgh, Edinburgh EH9 3JZ, United Kingdom }
\author{M.~Biagini}
\author{P.~Raimondi}
\affiliation{Laboratori Nazionali di Frascati dell'INFN, I-00044 Frascati, Italy }
\author{A.~Stocchi}
\affiliation{Laboratoire de l'Acc\'el\'erateur Lin\'eaire, F-91898 Orsay, France }
\author{S.~Bettarini}
\author{G.~Calderini}
\author{F.~Forti}
\author{M.~A.~Giorgi}
\author{A.~Lusiani}
\author{N.~Neri}
\affiliation{Universit\`a di Pisa, Dipartimento di Fisica, Scuola Normale Superiore and INFN, I-56127 Pisa, Italy }
\author{G.~Piredda}
\author{L.~Silvestrini}
\affiliation{Universit\`a di Roma La Sapienza, Dipartimento di Fisica and INFN, I-00185 Roma, Italy }
\author{M.~Ciuchini}
\affiliation{Universit\`a di Roma Tre, Dipartimento di Fisica and INFN, I-00146 Roma, Italy }
\author{Y.~Cai}
\author{S.~Ecklund}
\author{D.~W.~G.~S.~Leith}
\author{A.~Novokhatski}
\author{B.~N.Ratcliff}
\author{A.~Roodman}
\author{J.~Seeman}
\author{M.~Sullivan}
\author{U.~Wienands}
\author{W.~J.~Wisniewski}
\author{D.~B.~MacFarlane}
\affiliation{Stanford Linear Accelerator Center, Stanford, CA 94309, USA }
\author{F.~Martinez-Vidal}
\affiliation{IFIC, Universitat de Valencia-CSIC, E-46071 Valencia, Spain }
\author{T.~J.~Gershon}
\affiliation{Department of Physics, University of Warwick, Coventry CV4 7AL, United Kingdom }
\author{M.~Pierini}
\affiliation{University of Wisconsin, Madison, Wisconsin 53706, USA }

\fi

\begin{abstract}
This paper  is based on the outcome of the activity that has taken place 
during the recent workshop on ``\SuperB\ in Italy'' held in Frascati on November 
11-12, 2005. The workshop was opened by a theoretical introduction of 
Marco Ciuchini and was structured in two working groups. One focused on 
the machine and the other on the detector and experimental
issues.$^*$

The present status on \CP\ is mainly based on the results achieved 
 by \babar\ and \Belle.
Estabilishment of the indirect \CP\ violation in \B\ sector in 2001 
and of the direct CP violation in 2004 
thanks to the success of \pep2\ and KEKB \epem\ asymmetric 
$B$ Factories operating at the center of mass
energy corresponding to the mass of the \FourS.  
With the two $B$ Factories taking data, the Unitarity Triangle is now beginning to be 
overconstrained by improving the measurements of the sides and now also of the angles $\alpha$, and
$\gamma$. We are also in presence of the very intriguing results about
the measurements of \stwob\ in the time dependent analysis of decay
channels via penguin loops, where  $\b \to \s\sbar\s$ and  $\b \to 
\s\dbar\d$. \mtau\ physics, in particular LFV search, as well as charm and ISR physics 
are important parts of the scientific program of a \SuperB Factory.
The physics case together with possible scenarios for
the high luminosity \SuperB Factory based on the concepts of the
Linear Collider and the related experimental issues are discussed.
\vspace{1ex}

\null$^*${\footnotesize
 PARTICIPANTS: Justin Albert,
 David Alesini,
 Virginia Azzolini,
 Rinaldo Baldini Ferroli,
 Marica Biagini,
 Caterina Biscari,
 Yunhai Cai,
 Giovanni Calderini,
 Massimo Carpinelli,
 Marco Ciuchini,
 Alessia D'Orazio,
 Riccardo De Sangro,
 Emanuele Di Marco,
 Riccardo Faccini,
 Giuseppe Finocchiaro,
 Francesco Forti,
 Yoshihiro Funakoshi,
 Alessandro Gallo,
 Tim Gershon,
 Marcello Giorgi,
 Susanna Guiducci,
 David Hitlin,
 Toru Iijima,
 David Leith,
 Eugene Levichev,
 Luigi Li Gioi,
 David MacFarlane,
 Fernando Martinez Vidal,
 Nicola Neri,
 Sergey Nikitin,
 Fernando Palombo,
 Ida Peruzzi,
 Marcello Piccolo,
 Maurizio Pierini,
 Pavel Piminov,
 Giancarlo Piredda,
 Steve Playfer,
 Andrea Preger Miro,
 Pantaleo Raimondi,
 Aaron Roodman,
 Emmanuele Salvati,
 John Seeman,
 Dmitry Shatilov,
 Michael Sullivan,
 John Walsh,
 Andy Wolsky,
 Mikhail Zobov 
}
\end{abstract}

\maketitle

\newpage\null\newpage\null\newpage
\section*{\MakeUppercase{Outline of the document}}

This document is divided in three sections: 
\begin{itemize}
\item part one: physics motivations
\item part two: experimental issues from ``Detector Working Group''
\item part three: machine issues from ``Accelerator Working Group''
\end{itemize}

\part{\LARGE Overview and Physics Introduction}

\input{introduction}

\part{\LARGE Detector Working Group Report}

\input{detPart}

\part{\LARGE Initial Parameters for a Linear Super-B-Factory}

\input{accPart}

\input{biblioMerged}

\end{document}

%% file: introduction.tex

\section{Present \babar and \Belle performance}\label{subsec:lum}
The two existing asymmetric $B$ Factories, \pep2\ and
 KEKB, started their operations in 1999 and since then 
 their design peak luminosities have been exceeded:
 now \pep2\ is running with a 
peak luminosity of $ 10.5 \times 10^{33}$ $cm^{-2}$ $s^{-1}  $ and KEKB with
 $ 15.8 \times 10^{33}$ $cm^{-2}$ $s^{-1} $. 
These terrific luminosities have been obtained also thanks to the continuous beam
injection that both laboratories have assumed as normal operation mode.

\begin{table*} [bht]
\centering
\caption{\pep2 and KEKB design parameters.}\label{tab:pepkek}
\begin{tabular}{|c|c|c|} 
\hline 
\raisebox{0pt}[12pt][6pt]{Parameter} & 
\raisebox{0pt}[12pt][6pt] {\pep2}   & 
\raisebox{0pt}[12pt][6pt]{KEKB} \\ 
\hline
\raisebox{0pt}[12pt][6pt]{Peak Lumi} &
\raisebox{0pt}[12pt][6pt]{$ 3.0 \times 10^{33} \cms $}  &
\raisebox{0pt}[12pt][6pt]{$ 1.0 \times 10^{34} \cms$}  \\
\hline
\raisebox{0pt}[12pt][6pt]{$e^+$ Energy}& 
\raisebox{0pt}[12pt][6pt]{$9.0\gev$} & 
\raisebox{0pt}[12pt][6pt]{$8.0\gev$} \\  
\hline
\raisebox{0pt}[12pt][6pt]{$e^-$ Energy}& 
\raisebox{0pt}[12pt][6pt]{$3.1\gev$} &
\raisebox{0pt}[12pt][6pt]{$3.5\gev$}\\\hline 
\end{tabular}

\end{table*}
 
\begin{figure*}[cht]
\begin{center}
\includegraphics[width=0.70\textwidth]{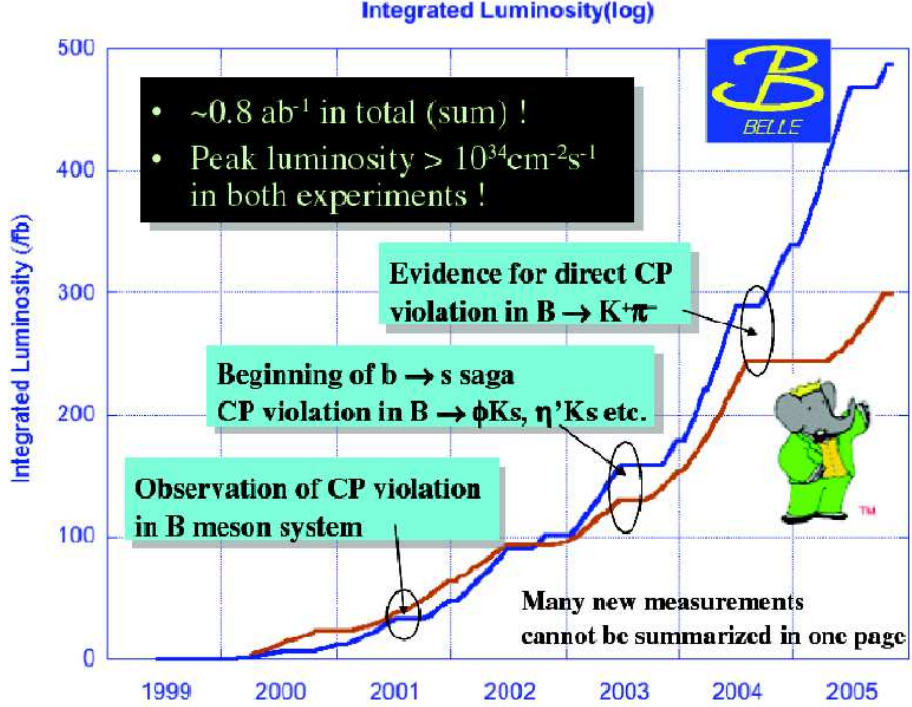} 
\caption{Integrated luminosities delivered by \pep2\ and by KEKB. } 
 \label{fig:lumi}
\end{center}
\end{figure*}

Both machines achieved a tremendous increase in the integrated
luminosity soon after the beginning of the operations in 1999 when 
both machines started to work in factory mode. 
Their peak luminosities exceed soon $10^{33}$ $cm^{-2}$ $s^{-1}$ , 
and the integrated luminosities of both $B$ Factories doubled every 2 years
as shown in figure ~(\ref{fig:lumi}). At present the total integrated
luminosity recorded by the two experiments 
(330 $fb^{-1}$ of \babar\ and the 500 $fb^{-1}$ of \Belle) is not
too far from one  billion of $B {\bar{B}}$ pairs.

\section{Detectors characteristics}
\label{subsec:det}

Both \babar\ and  \Belle detectors are conceptually similar, 
they appear at first glance as the typical hermetic apparata designed for \epem\ 
colliders, made of an inner Vertex Detector, a Tracking system, an 
Electromagnetic Calorimeter, a Solenoidal Magnet and a Muon/Hadron system.
Actually they show a clear asymmetry that is a consequence of
the machine energy asymmetry. The distribution of the decay products in the laboratory 
system is in fact peaked in the forward direction, the direction of the high energy beam.

The more relevant differences of \Belle with respect to \babar\ are in the Silicon Vertex
Tracker (a lower number of layers), in the Cherenkov system (Aerogel 
instead of the imaging internal reflection quartz DIRC) 
and glass instead of bakelite RPC for muon detection.


\begin{figure*}[ht]
\includegraphics[width=0.45\textwidth]{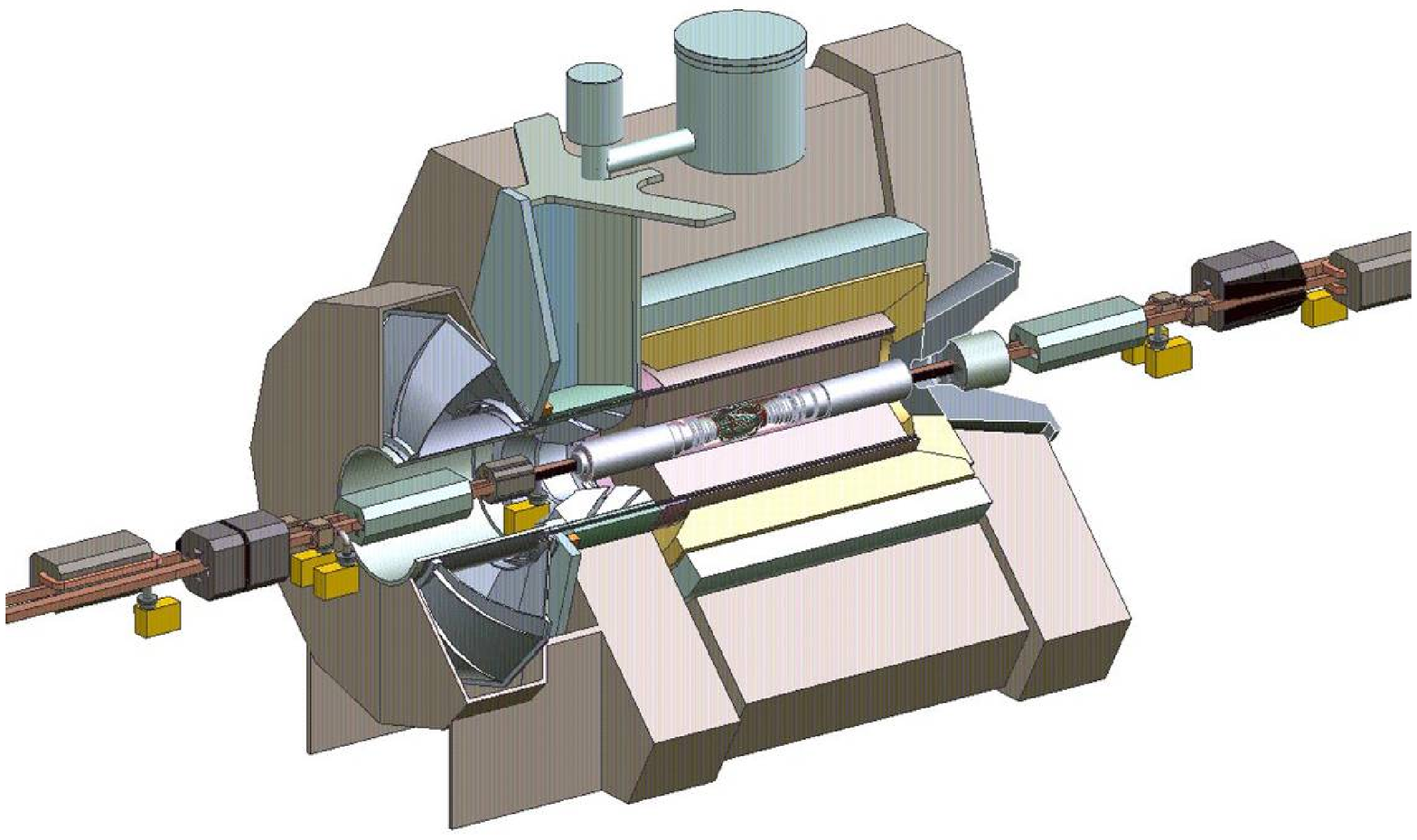}
\includegraphics[width=0.45\textwidth]{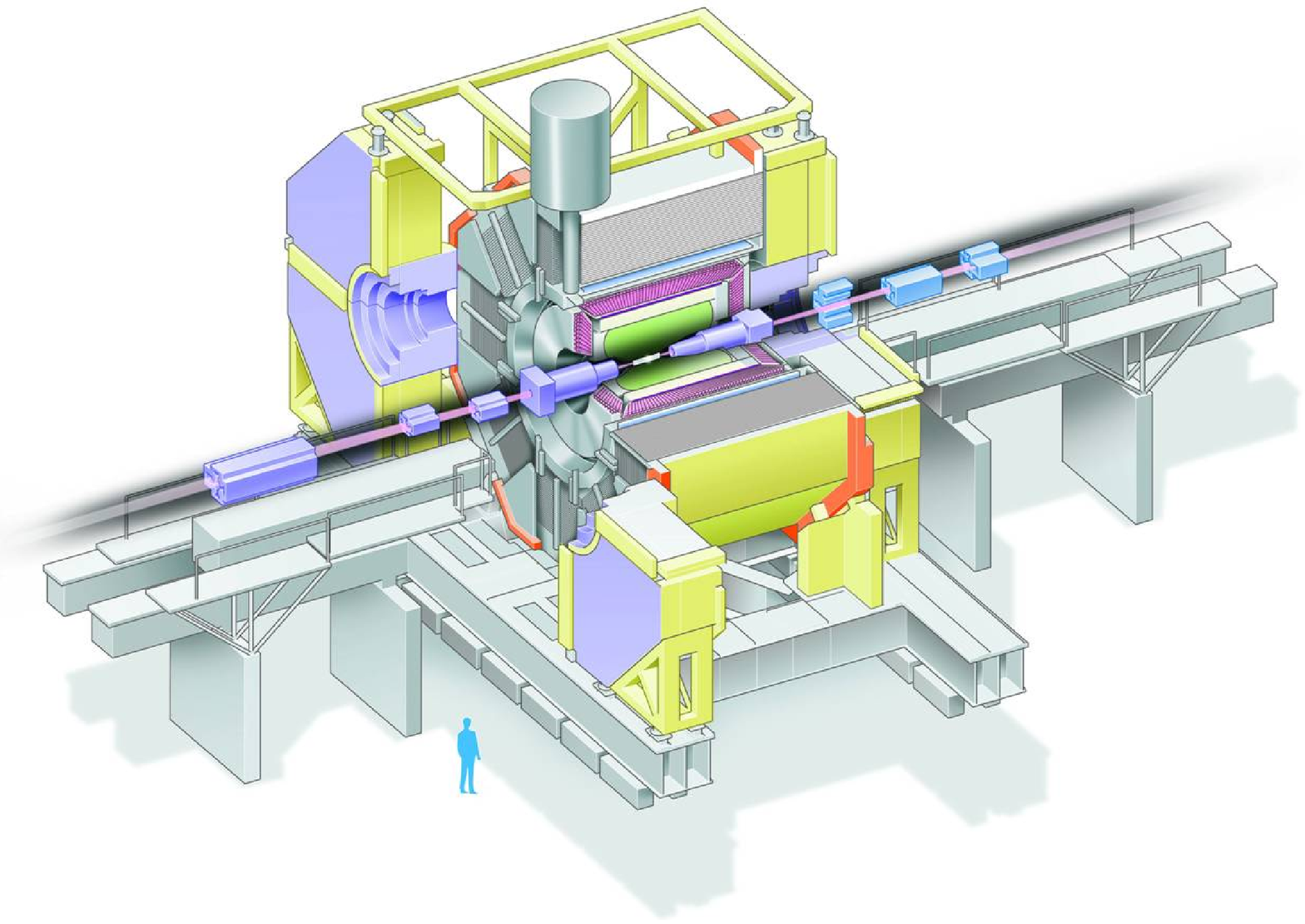}
\caption{\babar\ and \Belle detectors}
\label{fig:det}
\end{figure*}

\section{CKM constraints from new measurements of the Sides and Angles}
\begin{figure*}
\begin{center}
\includegraphics[width=0.40\textwidth]{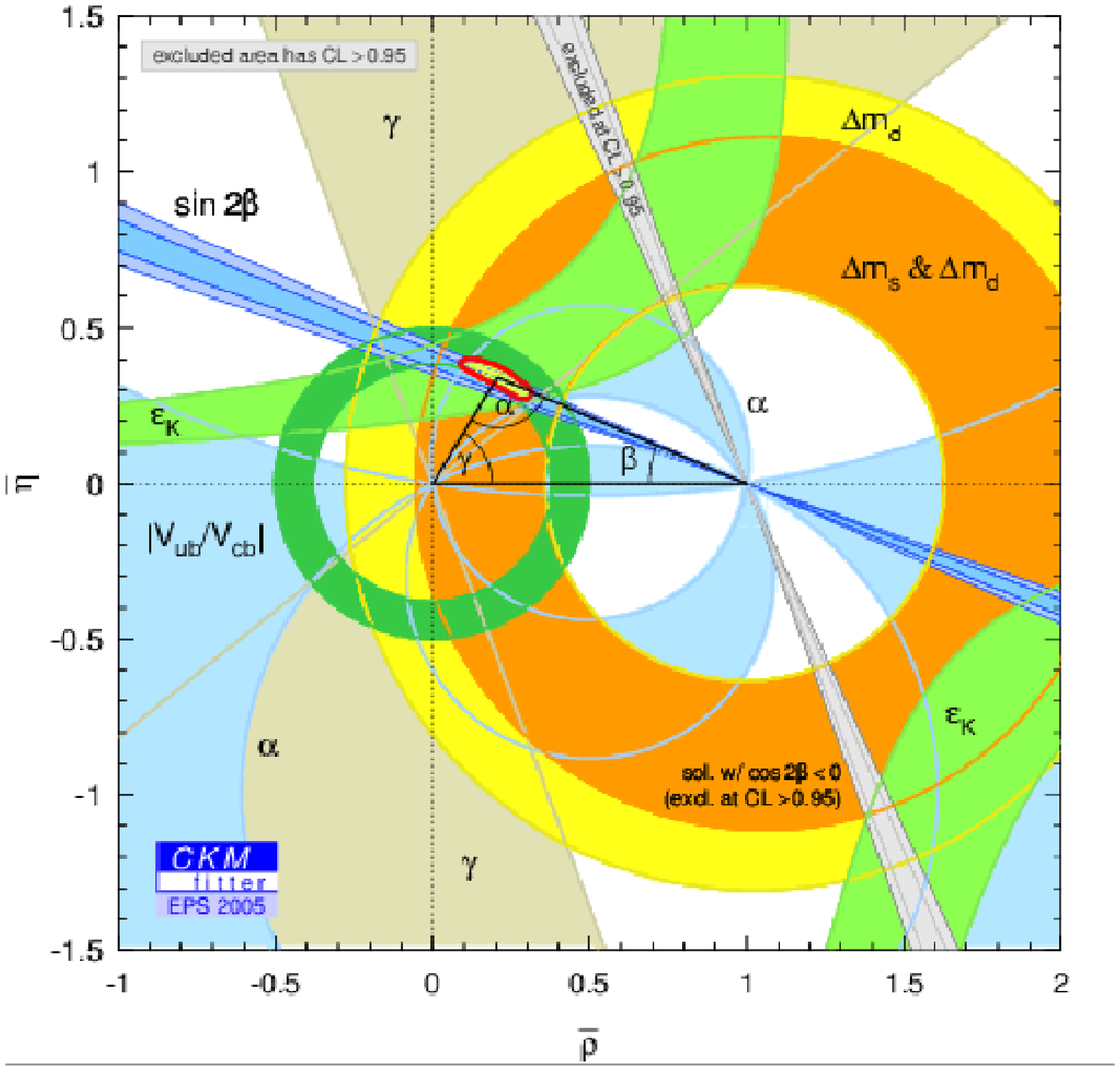}
\includegraphics[width=0.45\textwidth]{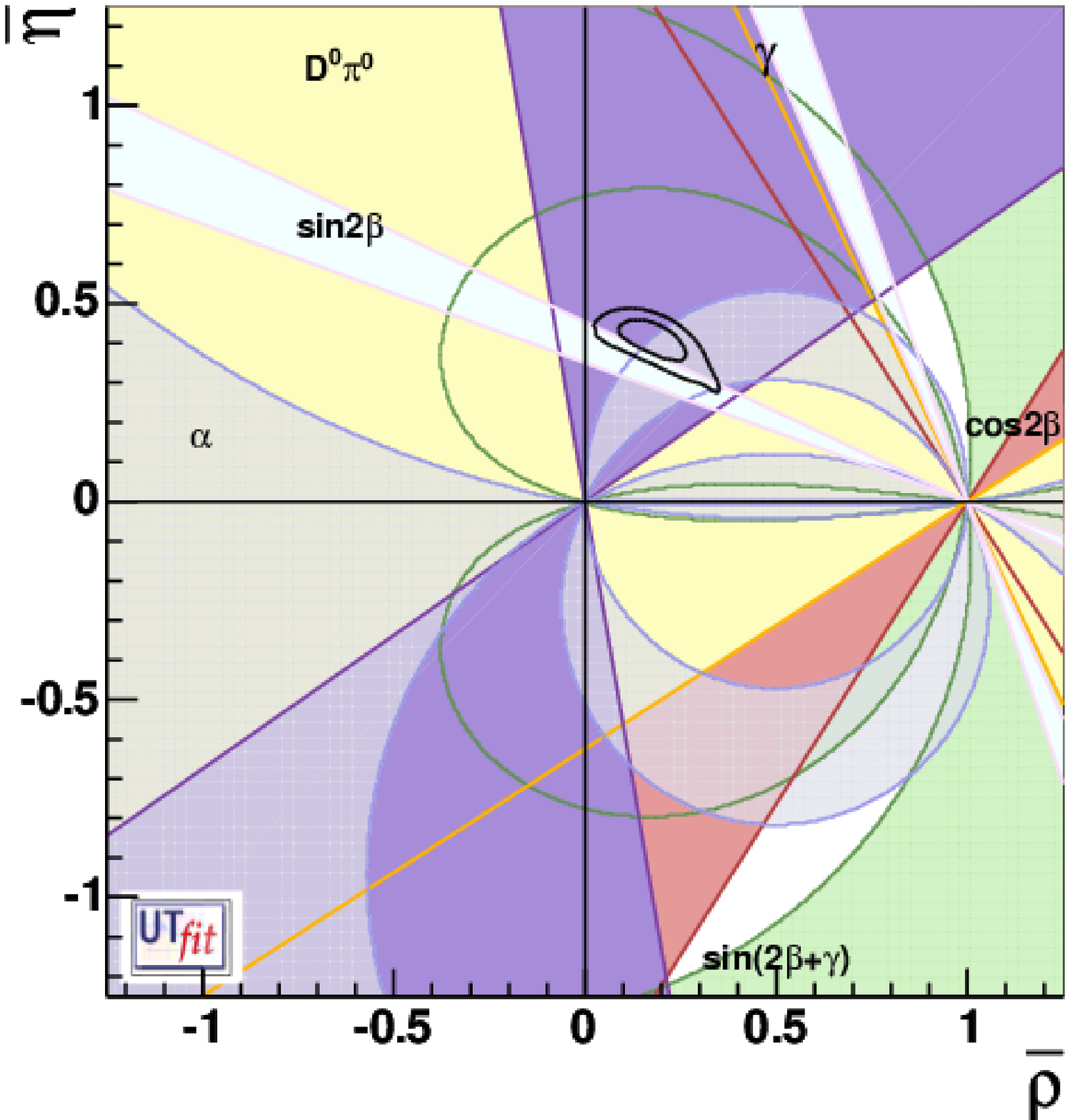}
\caption{ ${\bar{\rho}}$, ${\bar{\eta}}$  plane using all the available constraints (left) and only the information
from the UT angles (right).}
\label{fig:ckm1}
\end{center}
\end{figure*}

The measurement of \stwob\ became a program of precision
measurements in the general fits of the CKM matrix in the
$(\bar \rho , \bar \eta)$ plane already in 2004~\cite{sin2b-exp}. More challenging were the 
extractions of the other two angles of the unitarity triangle $ \gamma $ and
$\alpha$.
New values of $|V_{ub}|$ and $|V_{cb}|$~\cite{semilep}, a new value of $\stwob$ from charmonium~\cite{news2b}, 
the measurements of $\alpha$~\cite{alphaexp} and $\gamma$~\cite{gammaexp} have been presented last summer at LP05 and EPS05. 
They are in fact becoming the most stringent constraints of the Unitarity Triangle
in the $(\bar{\rho}, \bar{\eta})$  plane~\cite{fits}, as shown in the figure~(\ref{fig:ckm1}).
For this plot the information on $\cos 2 \beta $, $\Delta m_d$, $\Delta m_s$ and the direct CP
violation in the kaon sector $\epsilon_K$ are also used.
We also show the impact of the measurements of the angles from the $B$ factories in the right plot of 
figure~(\ref{fig:ckm1}). The agreement between this bounds 
and the global area coming from sides measurements is an important test of the consistency of 
the CKM mechanism in describing non-leptonic $B$ decays and CP asymmetries. Present data suggest a (not
yet significative) discrepancy, which calls for more data to be clarified in details.

\begin{figure*}
\begin{center}
\includegraphics[width=0.46\textwidth]{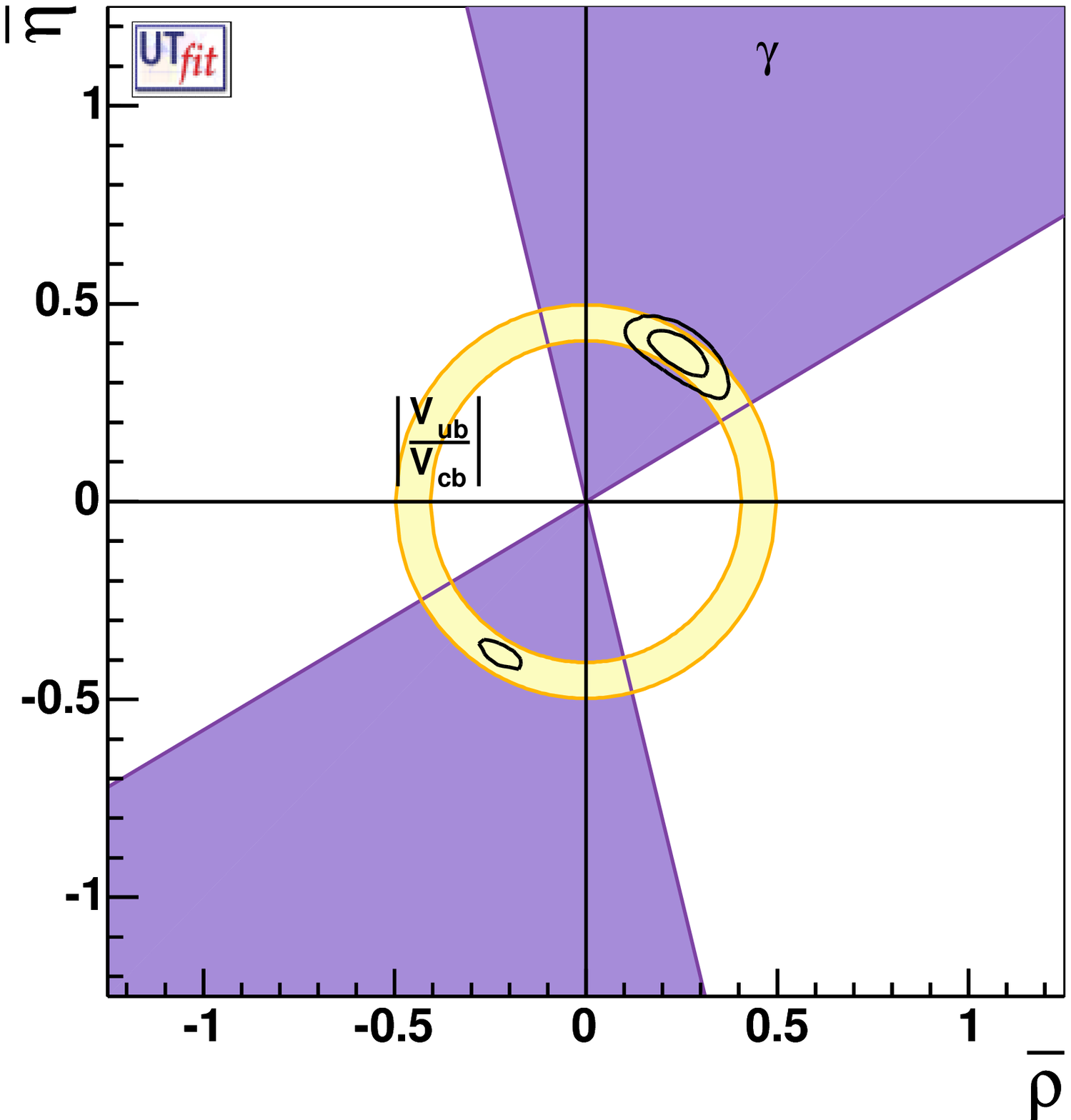}
\includegraphics[width=0.47\textwidth]{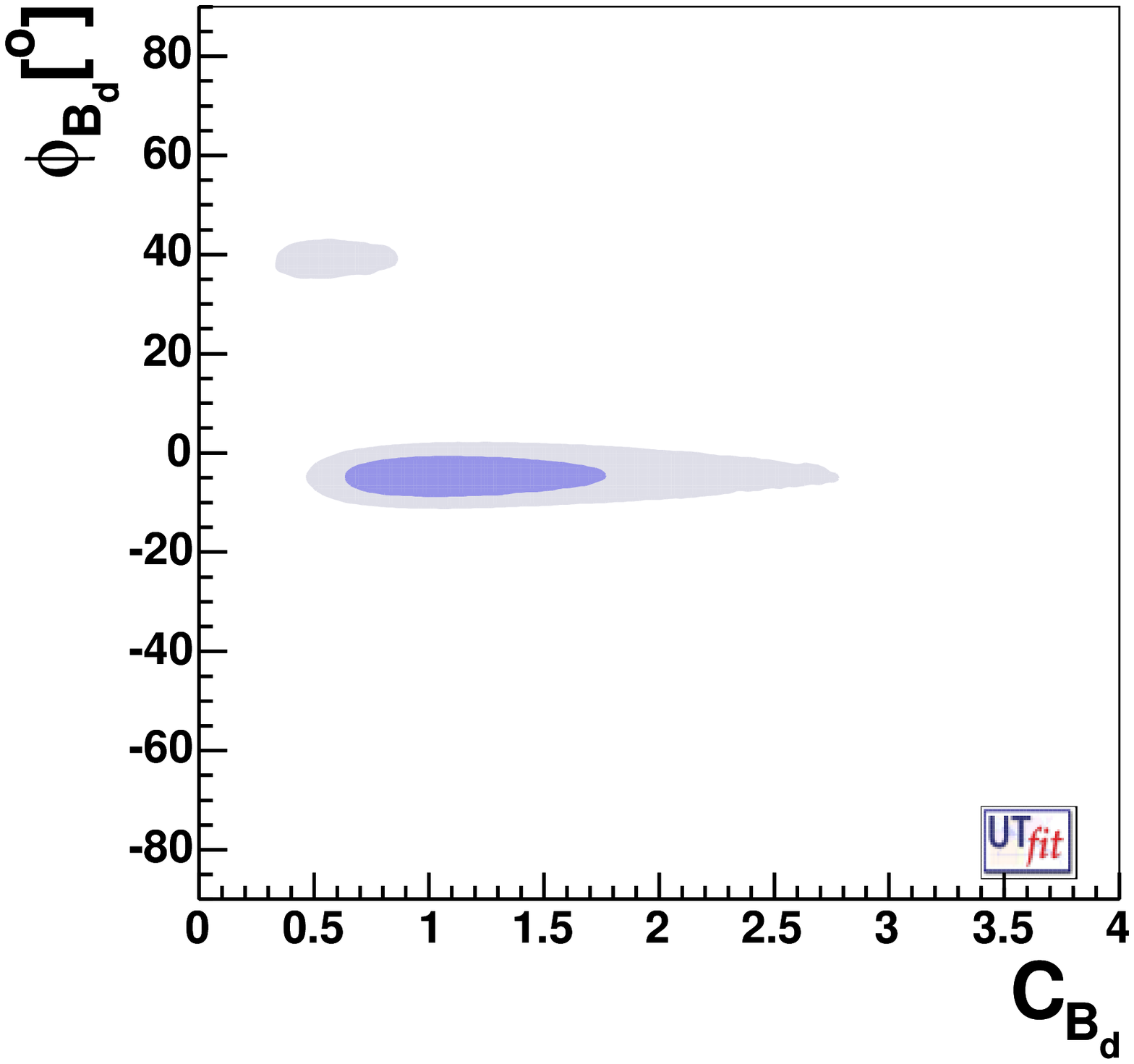}
\caption{Left:${\bar{\rho}}$, ${\bar{\eta}}$  plane in a NP generalized scenario.
Right: corresponding bound on NP parameters  $\phi_{B_d}$ and $C_{B_d}$ 
(see text)}
\label{fig:utNP}
\end{center}
\end{figure*}

 Making the hypothesis that NP enters observables in the flavour sector 
only at the loop level, it is possible to determine the $\bar \rho-\bar \eta$ 
plane independently of NP contributions (see Figure \ref{fig:utNP}), using tree-level $B$ decays :  
$V_{ub}$ and $V_{cb}$ using semileptonic inclusive and exclusive $B$ 
decays and the angle $\gamma$ measuring the phase of $V_{ub}$ appearing 
in the interference between $b \to c$ and $b \to u$ transitions to $DK$ final states.

The abundance of information allows to generalize the UT fit beyond the
Standard Model, parameterizing in a general way the effect of New Physics (NP) through
a multiplicative factor to the amplitude ($C_{B_d}$) and an additional
weak phase ($\phi_{B_d}$) in the $B$--$\bar B$ mixing process~\cite{NPUT}.
 As the left plot of figure~(\ref{fig:utNP}) shows, even in this case the 
present measurements give a good constrain on ${\bar{\rho}}$ and ${\bar{\eta}}$, strongly suppressing
the possibility of large NP enhancements in this sector (the fit gives a fraction of 7{\%} 
probability for the NP solution with negative ${\bar{\eta}}$, respect to the
93{\%} of the ``Standard Model like'' solution)~\cite{UTNP}. This means that a huge increase in luminosity
is needed, in order to discrimate NP scenarios from a simple Standard Model fit. This is more
evident on the right plot of figure~(\ref{fig:utNP}), where the present bound on
the $\phi_{B_d}~vs.~C_{B_d}$ plane is shown. In this case, in fact, the level of precision 
is not good enough to obtain any exclusion of the Standard Model scenario ($C_{B_d}=1$ and
$\phi_{B_d}=0$).

In Summer 2004, the first observation of the asymmetry

 \[ A_{CP}= \frac{ N(\bar B^0 \to K^- \pi^+) -
 N( B^0 \to K^+ \pi^-)} {  N(\bar B^0 \to K^- \pi^+) +
 N( B^0 \to K^+ \pi^-)} \] 

gave the evidence of direct CP violation in the $B$ sector. 
\babar\ found an asymmetry value of 
\[A_{CP}= -0.133 \pm 0.030 _{Stat} \pm 0.009_{Syst}. \] 
The value, after the average of \babar{}, \Belle, and CDF results,  
is~\cite{Akp} \[A_{CP}= -0.114 \pm 0.020.  \]

Starting in 2005 we have moved from the \CP\ Violation discovery era 
(given by the asymmetries from time dependent and time integrated analyses as in the case of 
the direct CP violation in $\Bz \to K^+\pi^-$ , or in $B^+$ 
analysis for the extraction of $\gamma$), to a new era of precision measurements and 
constraints on the Unitarity Triangle.

\section{Measurements of $\stwob$ via penguin-mediated decays}

An interesting field under study is the time-dependent analysis of
those decay channels that can only proceed through $"penguin"$ diagrams,
such as the $b \to (s \bar s s)$ processes:
%
\begin{figure*}
\begin{center}
\includegraphics[width=0.90\textwidth]{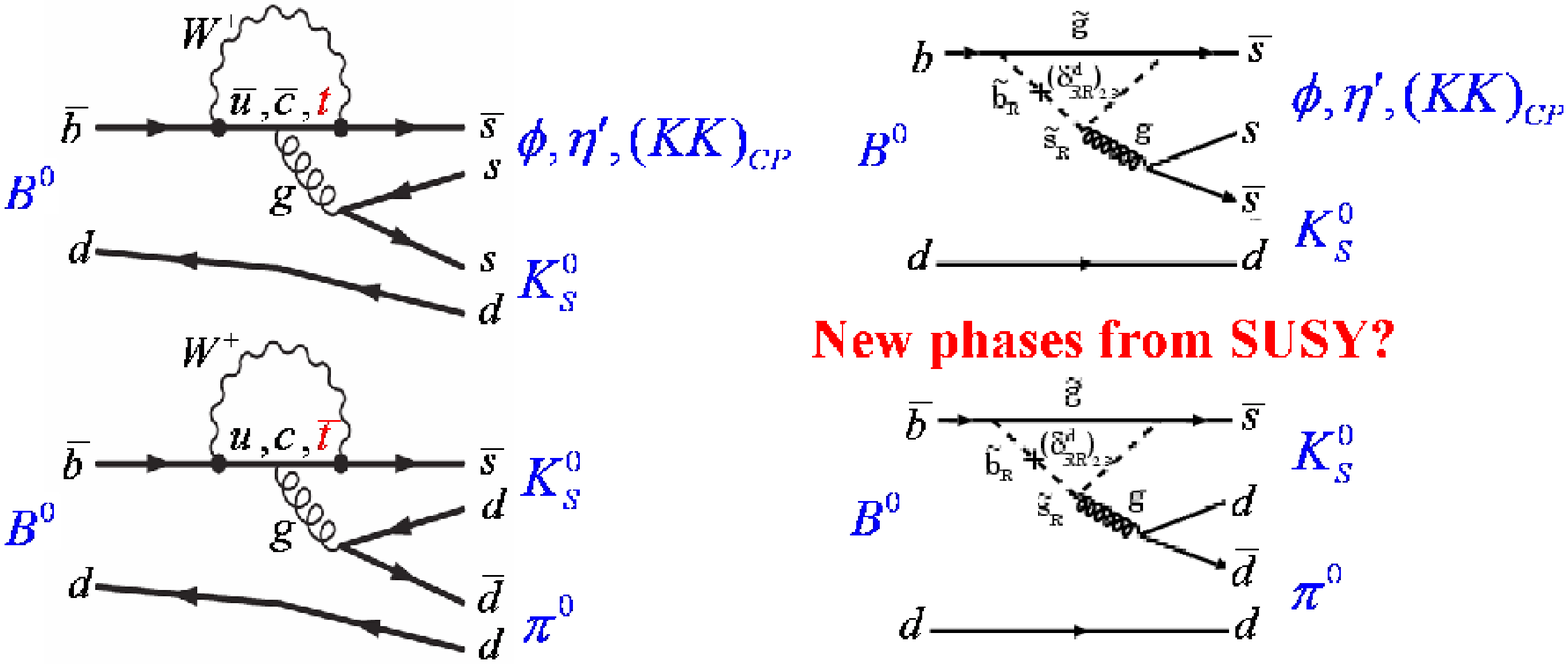}
\caption{Penguin diagrams.}
\label{fig:betapenguin}
\end{center}
\end{figure*}
\begin{itemize}
\item
$ B^0 \to \phi K^0$
\item
$ B^0 \to (KK)_{CP} K^0 $
\end{itemize}
and the similar $(b \to (d \bar d s))$ transitions:
\begin{itemize} 
\item
$ B^0 \to \eta \prime K^0$
\item
$ B^0 \to f_0 K^0$
\item
$ B^0 \to \pi^0  K^0$
\item
$ B^0 \to \rho^0 K^0$
\item
$ B^0 \to \omega K^0$.
\item
$ B^0 \to \pi^0  \pi^0 K^0$
\end{itemize}
\begin{figure*}
\centering
\includegraphics[width=0.45\textwidth]{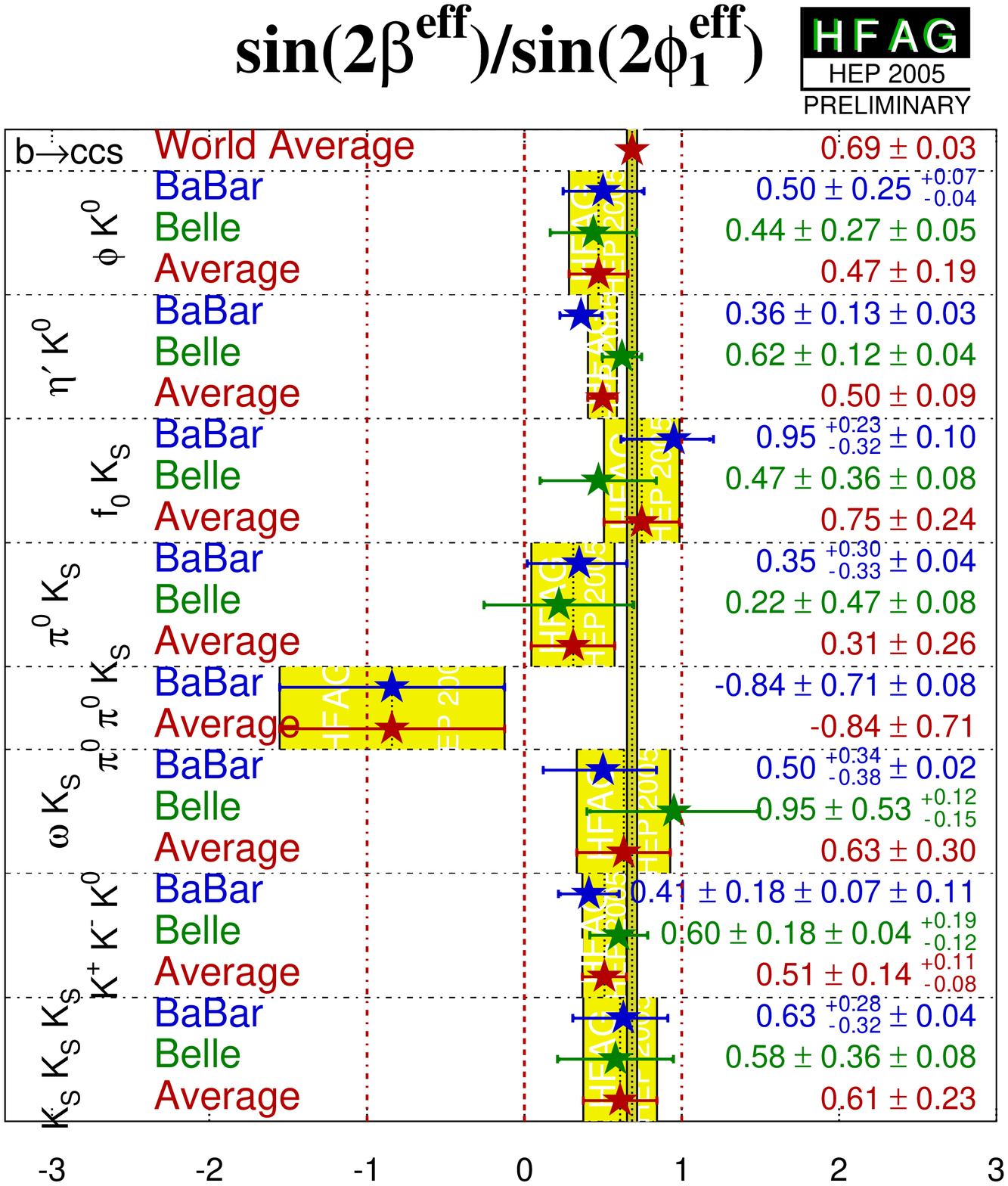}
\caption{Data for \stwob\ effective, comparing $\b \to \c\cbar\s$ and $\b \to \s\sbar\s$ 
$(\b \to \d\dbar\s)$.}
\label{fig:hfag}
\end{figure*}

These decays take the dominant contribution from the combination of CKM elements $V_{tb}V^*_{ts}$
and have the same phase of the charmonium channels  $b \to (c \bar c s)$, up to
a small phase shift of $V_{ts}$ respect to $V_{cb}$.
If, however, new heavy quanta contribute to the loops (as shown in figure~(\ref{fig:betapenguin})
in the case of SUSY), new phases can contribute to the asymmetry and the $S$ coefficient 
of the time dependent analysis could be substantially different from \stwob~\cite{NPbs}.
The comparison between results from all the above $B^0$ decay channels~\cite{bsexp} 
 and \stwob\ from charmonium, as shown in the HFAG plot~(see Figure {\ref{fig:hfag}}),
must be interpreted anyway $"cum$ $grano$ $salis"$. There are in fact other 
contributions to be taken into account in 
addition to the  diagrams with top quark insertion in the loop.

For example, even in the case the theoretically cleanest channels ($B^0 \to \phi K^0$ and
$ B^0 \to K^0_SK^0_SK^0_S $), one has to take into account a Standard Model uncertainty due to a penguin
contribution with an \emph{up} quark running in the loop. Using the CKM couplings to
scale this term to the leading contribution, we obtain a correction
of the order of $\lambda ^2$ $\approx 5 \%$ coming from the fact that these contributions are
doubly Cabibbo suppressed.
For the other decay channels the uncertainty could be as large as 
$10 \%$ (or even more), since in that case the doubly Cabibbo suppressed terms
also include tree-level transitions.~\cite{bstheory}

\begin{figure*}
\begin{center}
\includegraphics[width=0.40\textwidth]{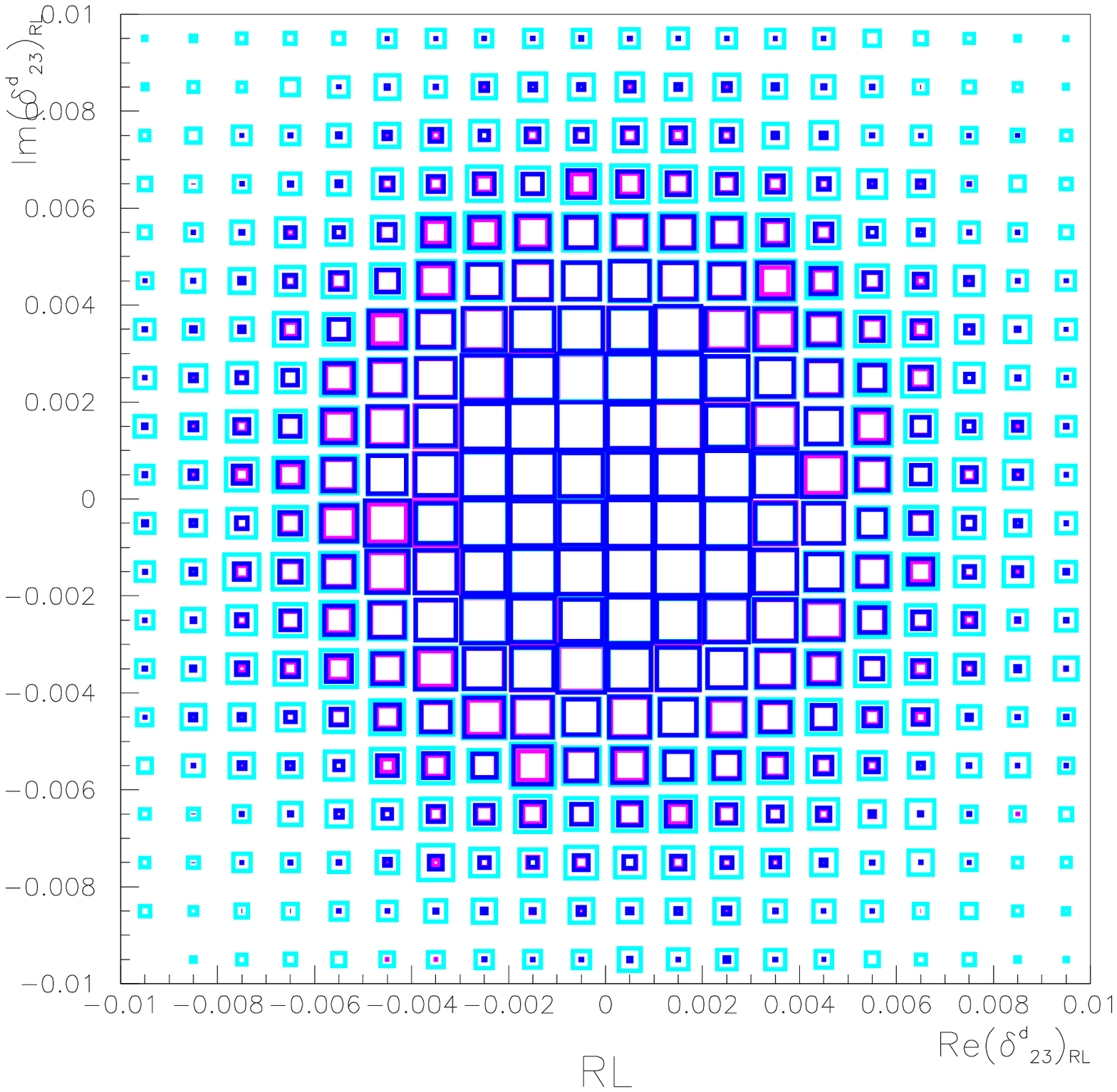}
\includegraphics[width=0.40\textwidth]{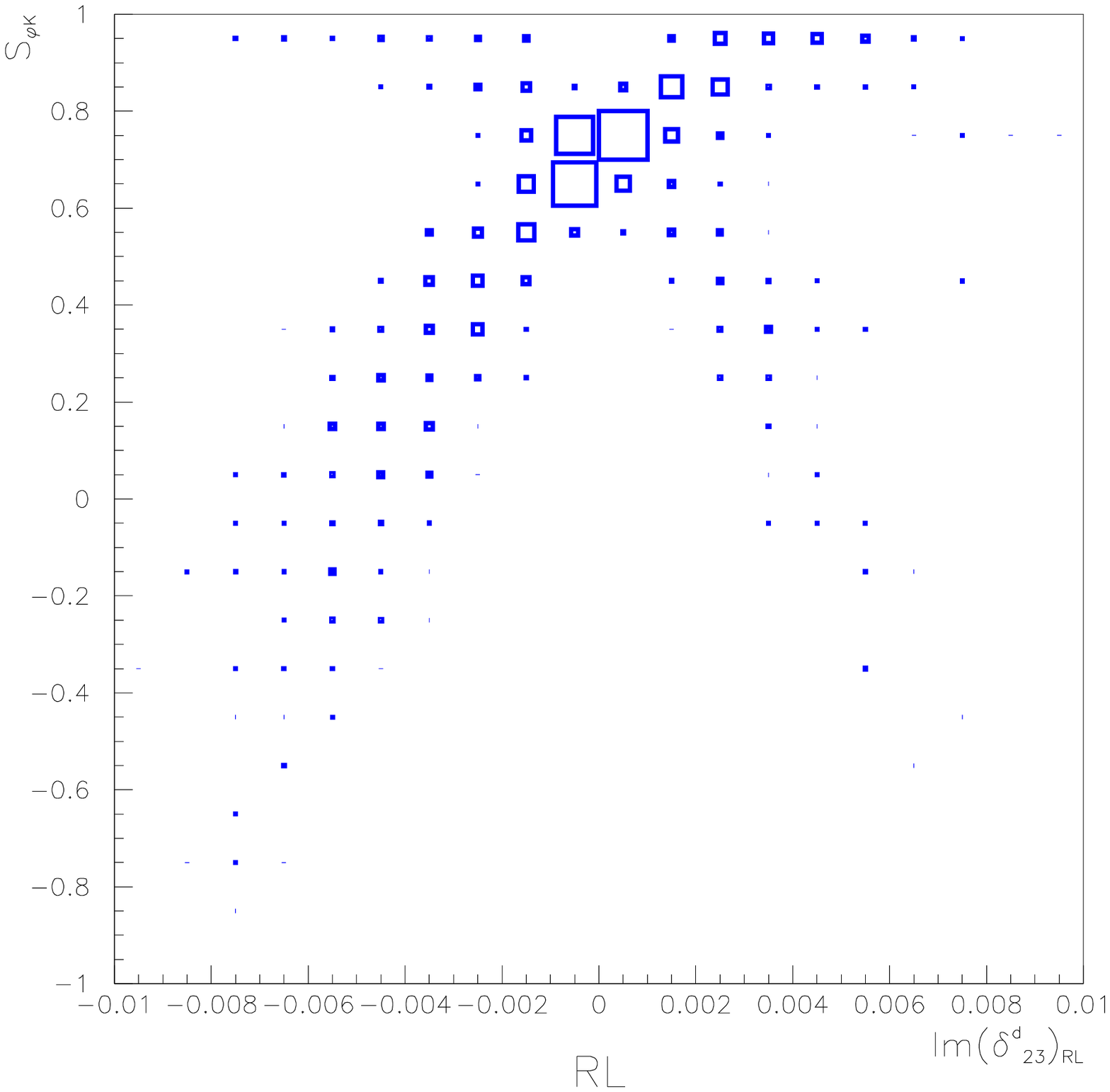}
\caption{Left: present bound on $Im(\delta^{23}_{RL})$vs.$Re(\delta^{23}_{RL})$ from $BR(b \to s \gamma)$ 
and $BR(b \to s ll)$.
Right: $S(\phi K^0)$ as a function of $Im(\delta^{23}_{RL})$ using the previous bound.}
\label{fig:masiero}
\end{center}
\end{figure*}

Once these contributions are taken into account, one can use the experimental results
for the $S$ parameters to obtain a bound on NP parameters~\cite{bsNP}. For example, one can use the knowledge of
$b \to s \gamma$ and $b \to s ll$ Branching Ratios in SUSY models to bound the values of
NP parameters and study their effect on the $b \to s$ penguin modes~\cite{bsMassIns}.
This is done in figure~(\ref{fig:masiero}), where the case of a $RL$ mass insertion in the squark
propagator (inducing a transition from $\tilde{b}\to \tilde{s}$ transition) is taken into 
account.\footnote{In SUSY models, the quark field rotation that generates the CKM
matrix in the Standard Model, diagonalizing the quark mass matrix, also acts on squark mass
matrix. Differently than for quark, squark mass matrix is not necessarily diagonalized
by this rotation. Mass insertions are complex parameters representing the residual
off-diagonal terms of the matrix.~\cite{MassIns}} 
The left plot of the figure shows the present knowledge 
on the $Im(\delta^{23}_{RL})$vs.$Re(\delta^{23}_{RL})$ from $BR(b \to s \gamma)$ and $BR(b \to s ll)$, while the
right plot gives $S(\phi K^0)$ as a function of $Im(\delta^{23}_{RL})$.

Many channels have been explored so far, some of them with poor statistics,
some other richer. It is clear that we are at the beginning of a very interesting season
when we can start probing the Standard Model in the flavour sector. 
In order to make this comparison really possible, the statistical error should be reduced 
below the theoretical uncertainties. To do that, a statistics between 20 and 50 times higher 
than present (i.e. 15 to 40 ab$^{-1}$) is needed. In few years, with LHC running and just before the
beginning of the ILC project, a program of precision measurement at a \SuperB\ with a capability of 
delivering more than a few tens of billions of \BB\ pairs, will be complementary to LHC physics. 
For example, the precise measurement of channels mediated by loop diagrams, both in $b \to s$ and $b \to d$
transitions, will allow to determine the couplings for New Physics contributions, such as
the mass insertion parameters $\delta^{23}$ and $\delta^{13}$ in SUSY scenarios.
For instance, a mass insertion $\delta^{23}$ with an imaginary part of $\sim2\%$, with an average squark
mass in the range $\sim350-450\ \gev$ can produce a deviation of $S(\phi\KS)$ of the order of
20\% respect to $S(\jpsi\KS-\phi\KS)$, as shown in the right plot of figure~(\ref{fig:masiero}).
In order to establish such 20\% difference at the $5\sigma$ level, {\it i.e.} 
measuring $A_{\CP}(\phi\KS)=0.60\pm0.03$, and assuming the current per event sensitivity, we need 
to integrate a statistics corresponding to 30 ab$^{-1}$.

Similar constraints on New Physics can be obtained studying similar channels. For instance,
the radiative penguin decays $b\to s\gamma$ provide a particularly clean environment.
Direct \CP\ violation in these decays is expected to be $\approx 0.5\%$ in the Standard Model, but could be enhanced
by New Physics contributions to the penguin loop. Recent inclusive and exclusive 
measurements are just beginning to constrain such contributions. The information they provide at this
point exclude the possibility of huge variations respect to the Standard Model expectations. On the other side,
because of the limited statistics, the possibility of observing an enhancement of an order of magnitude is still
open, but only a \SuperB\ factory can provide the needed statistics. It is also important to stress the fact
that CP measurements are statistics limited, and will continue to be so until at least 10~ab$^{-1}$. 
With larger samples it would be  interesting to measure the direct \CP\ asymmetry in $b\to d\gamma$ decays where the 
Standard Model prediction is -12\%.  \babar\ has also shown that it is feasible to measure time-dependent \CP\ 
violation in $\Bz \to K^{*0}(\ra \KS\piz)\gamma$. In the Standard Model the sine term of the time dependent
CP asymmetry in this channel is  suppressed respect to $\sin 2\beta$, being proportional to a factor related to
the helicity suppression of left-handed respect to right-handed photons. 
This measurement, which is sensitive to New Physics couplings with the opposite 
helicity, will continue to be statistics limited up to 50\invab. 
An alternative method of studying the photon polarization in $b\to s\gamma$ is the Dalitz plot distribution of 
the $K\pi\pi$ system in $\Bz \to K\pi\pi\gamma$, but this also requires a large statistics sample.    

\begin{table*} [htb]
\begin{center}
\caption [ ] {Measurement precision for \CP\ asymmetries in rare decays sensitive to New Physics.}
\label{tab:rarecp}
\begin{tabular}{|l|c|c|c|c|}\hline\hline
\multicolumn{2}{|l}{{\rule[-3mm]{0mm}{9mm}\large\bf $C\!PV$ in Rare Decays}} & \multicolumn{3}{|c|}{\bf \epem Precision }  \\ \hline\hline
 {\bf Measurement\ \ \ \ \ \ } & {\bf Goal} & {\bf 3/ab} & {\bf 10/ab} & {\bf 50/ab}  \\ \hline
$S(\Bz \to \phi\KS)$ & $\approx 5\%$  & 16\% &8.7\% &3.9\%  \\ \hline
$S(\Bz \to \eta^{\prime}\KS)$ & $\approx 5\%$ & 5.7\%& 3\%& 1\%\\ \hline
$S(\Bz \to \KS\piz)$ & & 8.2\%&5\% &4\% \\\hline
$S(\Bz \to \KS\piz\gamma)$ &  SM: $\approx 2\%$ & 11\%&6\% &4\% \\\hline
$A_{\CP}$ $(b \to s\gamma)$ & SM: $\approx 0.5\%$  & 1.0\% & 0.5\% & 0.5\% \\\hline
$A_{\CP}$ $(B \to K^{*}\gamma)$ & SM: $\approx 0.5\%$  & 0.6\% & 0.3\% & 0.3\% \\\hline
\end{tabular} 
\end{center}
\end{table*}

\subsection{Rare Decay Branching Fractions}

Many rare $B$ decay modes can potentially give access to physics 
beyond the Standard Model via measurements other than of \CP-violating asymmetries. 
Some examples of these modes are listed in Table~\ref{tab:rare}.  
Typically, these decays do not occur at tree level and consequently 
the rates are strongly suppressed in the Standard Model.  Substantial enhancements 
in the rates and/or variations in angular distributions of final state 
particles could result from the presence of new heavy
particles in loop diagrams, resulting in clear evidence of New Physics.  
Moreover, because the pattern of observable effects in highly 
model-dependent, measurements of several rare decay modes can provide 
information regarding the source of the New Physics.  

\begin{table*}[htb]
\begin{center}
\caption [ ] {Measurement precision for rare decays sensitive to New Physics.}
\label{tab:rare}
{\footnotesize
\begin{tabular}{|l|c|c|c|c|}\hline\hline
\multicolumn{2}{|l}{{\rule[-3mm]{0mm}{9mm}\large\bf Rare Decays}} & \multicolumn{3}{|c|}{\bf \epem Precision } \\ \hline\hline
 {\bf Measurement\ \ \ } & {\bf Goal} & {\bf 3/ab} & {\bf 10/ab} & {\bf 50/ab}  \\ \hline
$\Vtd/\Vts\sim\sqrt{\frac{{\cal{B}}(b\to d\gamma)}{{\cal{B}}(b\to s\gamma)}} $ & & 19\% &12\%  &5\%  \\ \hline
$\BR(B\to D^{*}\tau\nu)$& ${\cal{B}}=8\times 10^{-3}$ &10\% &5.6\% & 2.5\% \\ \hline
$\BR(B \to s\nu\bar{\nu})$  & 1 exclusive:  &$\sim 1\sigma$ &$>2\sigma$ &$> 4\sigma$ \\ 
 $\ \ \ \ \ \ (K^{-,0},K^{*-,0})$ &$\sim4\times 10^{-6}$& (per mode)&(per mode) &(per mode) \\ \hline
$\BR(B_d \to {\rm invisible})$ & & $<2\times 10^{-6}$&$<1\times 10^{-6}$ &$<4\times 10^{-7}$ \\\hline
$\BR(B_d \to \mu\mu)$ & $\sim8\times 10^{-11}$ &$<3 \times 10^{-8}$ &$<1.6 \times 10^{-8}$ &$<7 \times 10^{-9}$ \\\hline
$\BR(B_d \to \tau\tau)$ &  $\sim1\times 10^{-8}$ & $<10^{-3}$ & ${\mathcal O}(10^{-4})$ & ? \\\hline\hline
$\BR(\tau \to \mu\gamma)$ & & & $ 10^{-9}$ & $ 10^{-9}-10^{-10}$ \\\hline\hline
\end{tabular} 
}
\end{center} 
\end{table*}

The ratio of $b\to d\gamma$ to $b\to s\gamma$ decays is directly related 
to the ratio $V_{td}/V_{ts}$. It is interesting to measure this ratio in penguin processes as well 
as through $B_d/B_s$ mixing, since New Physics enters in different ways. 
The ratio of the exclusive decays $B\to\rho\gamma$ and $B\to K^*\gamma$ can be accurately measured, 
but the precision of the determination of  $V_{td}/V_{ts}$ is limited by theoretical uncertainties of $\approx 12\%$ in the ratio of the form factors. 
A measurement of the ratio of the inclusive decays does not suffer from this uncertainty, but is experimentally rather challenging, and requires a large data sample.

Searches for $B \to s \nu\bar{\nu}$, either inclusively or exclusively, are extremely difficult, due to the presence of the two final state neutrinos.  The required sensitivity can, however, be obtained using the recoil method, in which the signal mode (in this case the exclusive $B \to K \nu\bar{\nu}$ and $K^* \nu\bar{\nu}$ modes)
is sought in the recoil against a fully reconstructed hadronic $B$ decay.  Assuming Standard Model branching fractions, extrapolation of current analyses suggest that we would expect
 a signal of $~10$ events in each of the four modes ($K^{-,0},K^{*-,0}$) although with a substantial background, with 3 ab$^{-1}$ of data.  A statistically significant signal would emerge
 in the combination of modes with approximately 10 ab$^{-1}$ even using a simple cut-and-count analysis.

The decays $B_d \to \ell\ell$ ($\ell = e, \mu, \tau$) are somewhat less promising in the sense that it appears impossible to reach the predicted Standard Model branching
fractions even with more than 50\invab\ of data.  Moreover, $B_d \to \mu \mu$ is expected to be accessible at both LHC$b$ and $B$TeV, and these experiments
will also be able to access $B_s \to \mu \mu$, which is expected to provide a more stringent test of New Physics.  However, even 10\invab\ of data will improve the
existing limits on these modes by an order of magnitude, and an $\epem$ $B$ Factory does have the advantage of also being able to search for $B_d \to \epem$ and
the (extremely challenging) $B_d \to \tau^+\tau^-$ mode.

\subsection{$s\ellell$, $K\ellell$, $\Kstar\ellell$ Decays}

The exclusive $K^{(*)}\ellell$ and inclusive $s\ellell$ decays have been intensively studied theoretically, as they provide a potentially unique window on New Physics. For example, in the Standard Model, the forward/backward asymmetry $A_{FB}$ of the lepton pair has a zero at lepton pair mass $\hat{s}_0=0.14$ \gev. In extensions of the Standard Model, this zero may be approached from the opposite direction, or may be altogether absent. This region of lepton pair invariant mass represents only a small fraction of the allowed kinematic region of these rare decays, so a large data sample is required to make this measurement. The measurement of $A_{FB}$ can be done at hadronic experiments, but only in the exclusive modes involving muons. Theoretical predictions are typically more precise for inclusive processes, which can only be measured at a Super $B$ Factory.  It is very important to compare $A_{FB}$ in muon and electron modes, as this asymmetry can be changed by the presence of a charged Higgs. Table~\ref{tab:kll} summarizes the achievable measurement precision.

\begin{table*} [htb]
\begin{center}
\caption [ ] {Measurement precision for $s\ellell,\ K\ellell,\ \Kstar\ellell$ decays.}
\label{tab:kll}
\begin{tabular}{|l|c|c|c|}\hline\hline
\multicolumn{1}{|l}{{\rule[-3mm]{0mm}{9mm}\large\bf $s\ellell,\  K^{(*)}\ellell$ Decays}} & \multicolumn{3}{|c|}{\bf \epem Precision } \\ \hline\hline
 {\bf Measurement\ \ \ \ \ \ }  & {\bf 3/ab} & {\bf 10/ab} & {\bf 50/ab}  \\ \hline
$\BR(B \to K\mumu)/\BR(B\to K\epem)$  & $\sim8\%$ & $\sim4\%$&$\sim2\%$ \\ \hline
$A_{\CP}(B\to \Kstar\ellell)$ (all)  &$\sim6\%$ &$\sim3\%$ & $\sim1.5\%$ \\
$\phantom{A_{\CP}(B\to \Kstar\ellell)}$ (high mass)   &$\sim12\%$ &$\sim6\%$ \\ \hline
$A_{F\!B}(B\to \Kstar\ellell)$\ : $\hat{s}_0$  &$\sim20\%$ &$\sim9\%$ &$\sim9\%$  \\
$\phantom{A_{F\!B}(B\to \Kstar\ellell)}:\ A_{\CP}$ & & &  \\\hline
$A_{F\!B}(B\to s\ellell):$ $\hat{s}_0$  & $\sim27\%$ & $\sim15\%$ & $\sim7\%$ \\
$\phantom{A_{F\!B}(B\to \Kstar\ellell)}\!\!\!\!\!\!:\ C_9,\ C_{10}$ &$36-55\%$ &$20-30\%$ &$9-13\%$\\\hline\hline
\end{tabular} 
\end{center}
\end{table*}

\section{Extrapolation to 50$ab^{-1}$}
\begin{figure*}
\begin{center}
\includegraphics[width=0.60\textwidth]{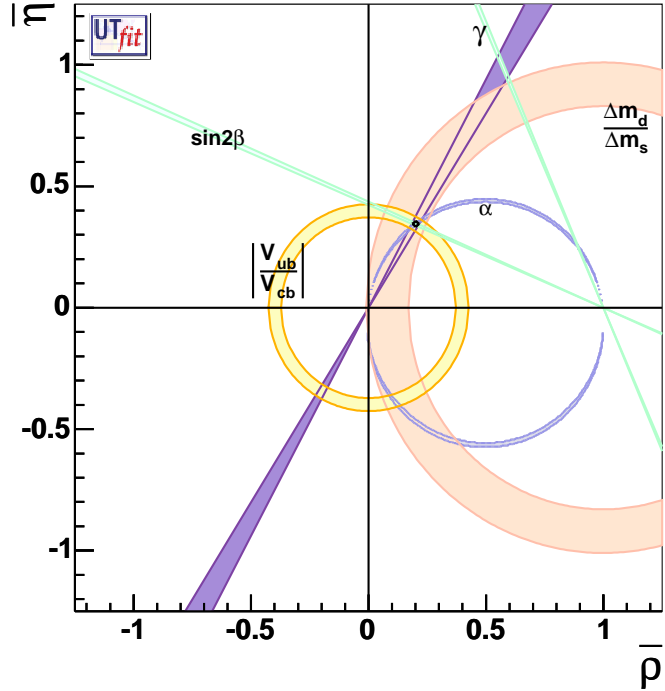}
\caption{Extrapolation of the Universal UT analysis to a statistics of $50ab^{-1}$.}
\label{fig:uut50ab}
\end{center}
\end{figure*}

The current experimental facilities are supposed to integrate a combined luminosity
of $\sim2ab^{-1}$, which will improve the present knowledge of SM related quantities and
will allow more stringent bounds on New Physics (NP) parameters. Never the less, the
precision will not be enough to significatively determine the values of such parameters,
even in the optimistic scenario of an early discovery of NP at LHC. A more
reasonable value for the statistics needed is $\sim 50 ab^{-1}$, which can be achieved in 
with a data taking of the order of one year at the facility proposed in this document.
With such a precision, the UT analysis will became a high precision test, as shown 
in fig.~\ref{fig:uut50ab}.

The plot represents the Universal Unitarity Triangle analysis~\cite{uut}, which is a UT fit 
performed using only quantities that are independent of NP contributions 
within MFV models.~\footnote{In practise, one cannot use $\epsilon_K$ and $\Delta m_d$ to
determine $\bar \rho$ and $\bar \eta$ independently of NP. On the other side,
after the UUT analysis is performed, these two bounds provide constraints on 
the scale of NP particles~\cite{UTNP,MFVisidori}.} This is a common starting point of the SM analysis, 
as well as of any study of MFV scenarios~\cite{BurasMFV}.
Even if this fit is obtained from the standard analysis removing those quantities
that are sensitive to NP (which provides a bound on NP parameters once the UUT is given as
input), we will be able to achieve a precision of the order of percent on $\bar \rho$ and $\bar \eta$.
This MFV generalized analysis will be precise enough to overcome the present SM fit, shown 
in figure~(\ref{fig:ckm1}).

\begin{figure*}
\begin{center}
\includegraphics[width=0.60\textwidth]{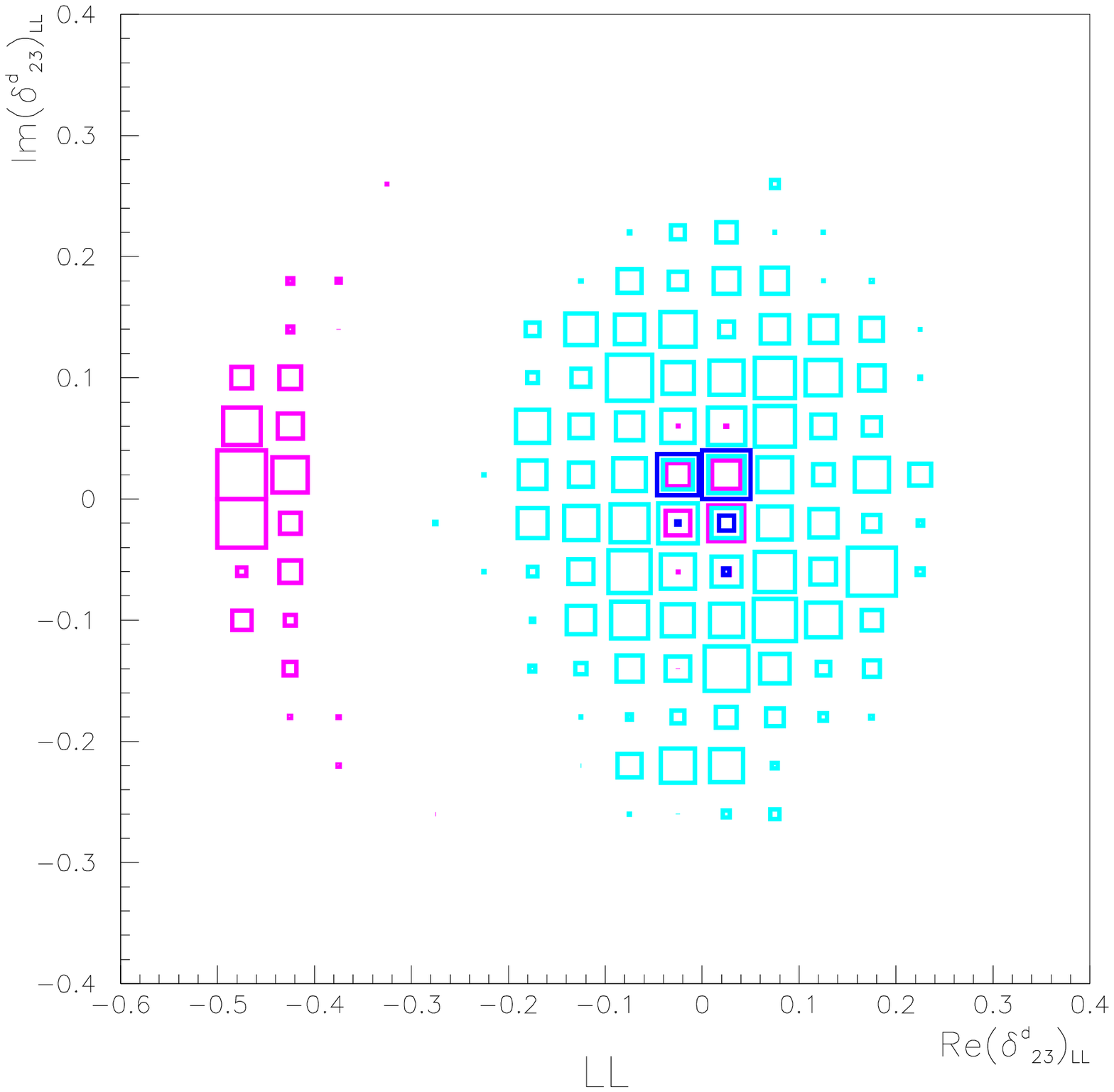}
\caption{Extrapolation of the the mass insertion analysis to $50ab^{-1}$:
bounds from $b \to s \gamma$(violet) and $b \to s ll$ (light blue) are shown, together
with the combined information (dark blue).}
\label{fig:masiero50ab}
\end{center}
\end{figure*}

At the same time, the improvement of the measurements of $b \to s$ processes will allow
to strongly bound the values of the mass insertions parameters of figure~(\ref{fig:masiero}),
as shown in figure~(\ref{fig:masiero50ab}). Here, we assumed the pessimistic scenario in which 
the two experimental inputs ($BR(b\to s\gamma)$ and $BR(b\to sll)$) will be in perfect agreement with the
SM. The experimental precision, in this case,  will be enough to test NP effects at the percentage
level. AT the contrary, there will be enough sensibility to translate any future experimental
discrepancy into useful information for the interplay between flavour physics and the direct
search of NP at the hadron colliders.

\begin{figure*}
\begin{center}
\includegraphics[width=0.80\textwidth]{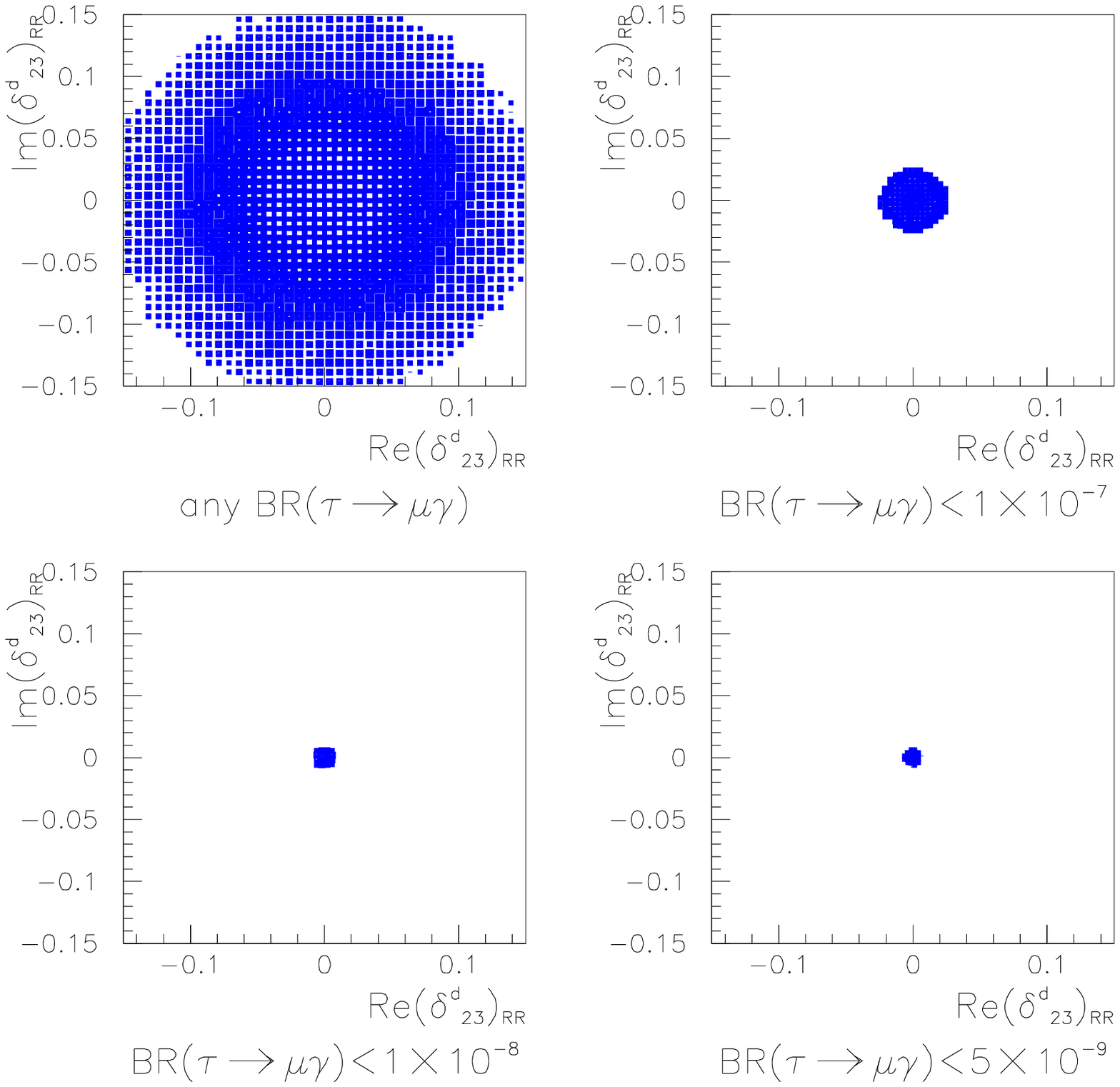}
\caption{Impact of the Upper Limit on $BR(\tau \to \mu \gamma)$ on the mass insertion parameters 
$\delta^{23}$ in GUT scenario.}
\label{fig:GUT}
\end{center}
\end{figure*}

Another important aspect to stress is the strong connection of $B$ and $\tau$ physics, in the 
framework of testing GUT models. An example of this is provided in figure~(\ref{fig:GUT}), where
the impact of the Upper Limit on $BR(\tau \to \mu \gamma)$ is shown on the same plot of figure~(\ref{fig:masiero}),
once $b \to s$ and $\tau \to \mu$ decays are connected in the framework of GUT.~\cite{LFV}
In general, with the high luminosity that a \SuperB\ factory can collect, rare decays of $\tau$ leptons
can be studied with high precision, providing a stringent test of flavour violation in the leptonic
sector and boosting our capability of testing GUT models.

 






%% file: detPart.tex

The \babar and \Belle detectors have proven to be very effective
instruments to explore precision flavor physics. The detector working
group considers that their basic design remains valid at \SuperB,
provided some subsystems can be modified as discussed in the
following. 

The \SuperB linear collider design employs low-current beams crossing
at a frequency in the MHz range with a very small beam spot size,
resulting in a relatively low machine background rate. In addition,
machine design and cost argument require a reduction in the energy
asymmetry changing the boost from the current $\beta\gamma=0.56$ of
\babar and 0.45 of \Belle down to $0.2-0.3$.

Under these conditions, in particular the low background rate, many of
the \babar or \Belle subsystems would be directly reusable at \SuperB,
or would require some design optimization within the same technology
choice, mainly driven by the beam crossing time structure.

On the other hand the smaller boost, reducing the separation between
the two $B$ decay vertices, requires an improved vertexing resolution
to maintain the physics reach. This improvement is possible thanks to
the small beam dimensions that allow a reduction of the beam pipe
radius to 1.0\cm or less, but will require significant R\&D on the
detector technology needed to instrument the small radius tracking
region. Particle identification will also require R\&D to improve the
compactness of the light detection system and extend angular
coverage. Some amount of R\&D will also be required to optimize
crystals for small angle calorimetry, where the high Bhabha rate will
increase occupancy beyond the capabilities of current CsI(Tl) systems.

In the following we discuss how machine parameters such as boost and
energy spread affect the detector design and physics reach
(Sec.~\ref{sec:depend}), what machine and bacgkrounds are expected
(Sec.~\ref{sec:backgrounds}), and the main issues for the detector
sub-systems (Sec.~\ref{sec:vertexing}~--~\ref{sec:muon})

\section{Physics dependance on basic parameters assumptions}
\label{sec:depend}
There are several parameters of the \SuperB Factory which directly
impact the physics capabilities at the new accelerator.  These
include: the energy asymmetry or $\beta\gamma$, the center-of-mass
energy spread, the beam size and beam-pipe size at the interaction
point, and the probability of multiple interactions in a single
crossing (or train of crossings).

The study of time-dependent \CP violation requires that the
$\Bz-\Bzb$ system be boosted in the lab system, so that the \B-meson
decay vertices are separated by a measurable amount.  The current \B
Factories have boosts of $\beta\gamma = 0.56$ (\pep2) and 0.45 (KEKB).
Lower boosts are feasible with improved vertex resolution, as
described below.

The $\Upsilon(4S)$ has a FWHM of roughly 18~\mev, so that the
accelerator's beam energy spread will reduce the effective
cross-section to $\BB$.  A center-of-mass energy spread of $5~\mev$
($10~\mev$) corresponds to a reduction of roughly 0.85 (0.67).
Exclusive reconstruction of \B-meson decays usually rely on two
kinematic variables to separate the signal from continuum (\qqbar)
or random combinatoric backgrounds.  Typically these kinematic
variables are $\mes = \sqrt{(E_{Beam}/2)^{2} - p_{B}^{2}}$ and $\Delta
E = E_{B} - E_{Beam}/2$, where all quantities are in the
center-of-mass.  The width of the $\mes$ distribution is determined
almost exclusively by the accelerator's center-of-mass energy spread,
and a larger energy spread reduces the background separation
achievable with this kinematic variable. Broadening of the $\mes$
distribution is ultimately limited by the width of the $\Upsilon(4S)$,
since the energy spread convoluted with the $\Upsilon(4S)$ line-shape
is the relevant distribution.  The $\mes$ width saturates at roughly
$5.5\mev$, or a factor of two wider than \pep2.

The much smaller beam-spots at the \SuperB machine will permit smaller
beam-pipes, allowing the first active layer of a Silicon Vertex
detector to be closer to the interaction point.  Such improvements
will be necessary to reduce the energy asymmetry, while maintaining
the ratio of \B vertex separation over vertex resolution at or above
$2.5$.  In addition, such improved vertexing may allow improved flavor
tagging and background rejection.  For example, it may be possible to
identify the charm vertex in semi-leptonic decays, to better separate
leptons from \B as opposed to charm decays.

Lastly, the time structure at the SuperB machine will be different
from that at the current \B factories. The bunch collision frequency
will be of order 1~MHz, compared to 238~MHz at \pep2.  At this
frequency, for a luminosity of $\lum = 10^{36}$, the roughly $100\nb$
cross-section, for all processes, corresponds to an interaction in
every tenth crossing.  This implies that in about 10\% of all \BB
events there will also be a Bhabha or two-photon interaction, and
roughly 1\% of the time there will be a second \qqbar or \BB event as
well.  This {\it physics pile-up} will produce additional background for
rare processes, but can be removed with global event energy
requirements.

\section{Machine Backgrounds}
\label{sec:backgrounds}

The reduction in beam currents in a linear-colliding $B$-Factory design,
as compared with a standard storage ring configuration, results in a dramatic
reduction in single-beam background (which would otherwise be the dominant background source).
The single-beam background sources, for example beam-gas, should be
negligible compared with the luminosity-scaling sources detailed below.
A precise background prediction
is not possible at this stage given the preliminary nature of the
machine design. However, the detailed characterization of the
various background sources on the present PEP-II and KEK-$B$ machines, as well as experience from SLC and studies for backgrounds at the ILC, allow us to
extrapolate the background sources which are important for consideration at present:
\begin{description}
\item[Luminosity Sources]
The most dangerous
background source is radiative Bhabha debris. In this case, the background
rate is strictly proportional to luminosity and therefore will be
100 times higher at a Super \B\ Factory than today. To control this source,
magnetic bending elements 
should be kept as far as is practical from the
interaction point in order to prevent off-momentum electrons or
positrons from reaching apertures close to the detector. At PEP-II, the presence of
a strong dipole field only 20\cm\ away from the IP, required to
separate the two beams in the absence of a crossing angle, accounts for the fact 
that this background is quite prominent while being small or negligible at KEK-\B.
At \SuperB, the first magnetic elements will be
focusing quadrupoles located approximately 40\cm\ away from the IP. 
Due to the lack of nearby dipoles,
the ``luminosity'' background should
be substantially reduced from a simple extrapolation from PEP-II, but potentially larger than indicated
by the KEK-\B extrapolation. We have assumed, as a baseline, a
factor 5 reduction with respect to the PEP-II extrapolation, 
but subdetectors have considered the full range between 0.2 and 1 
times the PEP-II-based extrapolation as a measure of the uncertainty in the estimates.
\end{description}
\begin{description}
\item[Beam-beam interactions.] 
Beam-beam interactions, {\it i.e.,} the transverse blowup of one beam due 
to its interaction with the other, will be one major source of 
particle loss rate. Backgrounds will be produced by beam tails which are
created by this mechanism and then hit nearby apertures. Quantitative prediction 
of these tails is almost impossible, especially  when the machine is operated 
close to its limits where they tend to get very large.  The best means to control 
the contribution is to design an advanced collimator scheme that can prevent beam tails 
from hitting apertures near the detector. The detector must also be well protected 
against the secondaries produced at the collimators. The maximum amount of collimation 
will be determined by the lifetime loss observed when closing the collimators. It is 
therefore prudent to integrate the collimator design as early as possible into 
the machine design to have control of this background source. 
The observed PEP-II beam-beam terms have been roughly parameterized 
and extrapolated, leading to a conclusion that this source will make a negligible contribution. 
However, it will require a very substantial effort to realize such performance at \SuperB.  
\item[Touschek background.] 
Touschek background, {\it i.e.,} longitudinal beam blowup due to intrabunch scattering,
will be present and is proportional to the bunch density. It will
therefore be significantly increased with respect to the present \pep2\ value.
This effect accounts for a significant part of the \Belle
background (around 20\%) but is barely visible in PEP-II. 
Its effect on the detector will again be a very sensitive
function of the magnetic elements in that area. This
background can be modeled in principle rather well and a good extrapolation should be 
available in the coming months.
\item[Synchrotron radiation.]
As noted above, we expect the \textit{single-beam} sources of synchrotron radiation to be negligible, due to
the low beam currents.  The amount of \textit{beam-beam} synchtrotron radiation (``beamstrahlung'') 
should be greater than either the present $B$-Factories or in a super-$B$ configuration 
using a standard storage ring; however it should be small compared with, for example, the ILC,
or even the SLC, due to the much lower beam energies.  A 5 $\mu$m gold masking similar to that
presently within the PEP-II beampipe should be sufficient to remove this contribution (and
may perhaps be able to be made even thinner).
\end{description}

All subdetectors are required to satisfy a factor of five safety margin to take into 
account for these extra terms.  A large safety margin is also needed to cope with 
background fluctuations.

To guarantee high operational efficiency, it is essential
that the detector can stand short background bursts, of magnitude comparable with the steady level.
Such bursts, typically lasting a few seconds, are seen very frequently (up to 1 per hour at KEK-B) 
and are attributed to dust particles attracted from the beam pipe to the center of the beam. 
This implies detectors with  large safety margins in terms of 
integrated dose and radiation bursts.

Finally, proper instrumentation must be integrated
from the start into the detector and machine designs, in order to measure the background, isolate its 
various sources and monitor its evolution in a continuous fashion.

In conclusion, although backgrounds are expected to be significantly smaller at a linearly-colliding
$B$-Factory than in a configuration using a standard storage ring, they are still of crucial import in the design of
the detector and interaction region to minimize the sources of backgrounds that are not reduced by
a reduction in single beam currents.  In particular, the design of the IR to minimize radiative-Bhabha background, and
the ability of detector elements to withstand short background bursts, are both of critical importance.

\section{Vertexing}
\label{sec:vertexing}

The \SuperB interaction region design  is characterized by the
small size of the transversal section of the beams, fraction of $\mum$ for $\sigma_{x}$ and tens of 
nm for $\sigma_{y}$. Therefore it will be possible to reduce the radial dimension of the beam-pipe tube 
up to $5-10$ mm radius, still preventing the beams to scatter into the tube within the detector 
coverage angle. The Be thickness of the tube will be reduced down to $200-300 \mum $, corresponding to
$(5.6-8.5) \cdot 10^{-4} X_0$, depending on the value of the radius of the beam pipe.
Another feature of the innovative SuperB design is the low current circulating inside 
the beam-pipe, tens of mA, in spite of the high luminosity. 
The cooling system for the beam-pipe will be not necessary and it has been removed in the design.
The reduced amount of radial material and the possibility to measure the first hit of the track 
very close to the production vertex will benefit the track parameters determination.

First Monte Carlo studies indicate precise determination of the
$B$ decay vertices at the level of $10-20 \mum$ and consequently 
on the $\Delta z$ separation along the beam axis among the two $B$ mesons, 
basic ingredients for the time dependent analysises.

The multiple scattering contribution to the resolution on the decay vertex  
is no more dominating since the amount of radial material is very much reduced with respect to 
the B-factory scenario. The intrinsic spatial resolution fixed by the pitch 
width of the silicon vertex detector will add a non-negligible contribution on the vertex measurements and 
therefore it will be important to minimize it as much as possible in a future vertex detector.

\begin{figure*}[thb]
  \begin{center}
  \includegraphics[width=0.9\textwidth]{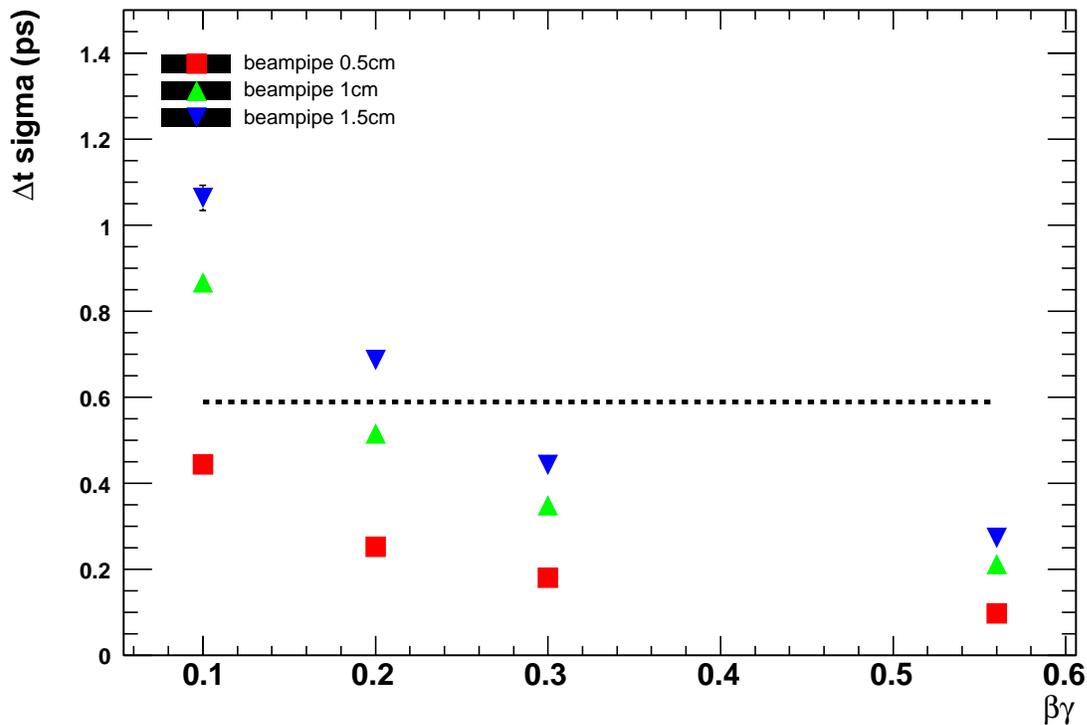}
  \caption{$\Delta t$ resolution as a function of the $\beta\gamma$ boost value of the center of 
mass rest frame for different beam-pipe configutions: 0.5 cm, 1.0 cm and 1.5 cm radius. The dashed line represents the \babar\ reference value. }
  \label{fig:DeltaT}
  \end{center}
\end{figure*}

For the simulation study we added a $50 \mum$ Monolithic Active Pixel silicon layer, in addition to the current \babar\ 
silicon vertex detector,  mounted on a $50 \mum$ kapton foil and glued directly on the beam-pipe.
This configuration allows to measure the first hit of the track just outside the beam-pipe.
In figure~\ref{fig:DeltaT}it is shown the resolution on the proper time difference of the two $B$ 
($\Delta t$) for three different beam-pipe configurations:

\begin{itemize}
\item{0.5 cm radius}: consider this as the aggressive scenario where we can evaluate the very limit of the 
vertex resolution we can think to achieve. We considered $200 \mum$ Be thickness for the beam-pipe and 
$5 \mum$ spatial resolution on hits from charged tracks.  
\item{1.0 cm radius}: most likely scenario with $300 \mum$ Be thickness for the beam-pipe and 
$10 \mum$ spatial resolution on hits from charged tracks.
\item{1.5 cm radius}: conservative scenario with $500 \mum$ Be thickness and 
$10 \mum$ spatial resolution on hits from charged tracks.
\end{itemize}

In all the configuration we have considered a $5-\mum$-thick gold foil
(equivalent to $150 \mum$ of silicon) before the first layer of the
vertex detector in order to absorb low energy background photons.

The spectacular precision of the decay vertex determination will also benefit several aspects of the 
$B$ meson reconstruction. The possibility to reduce the energy asymmetry without affecting the 
proper time resolution will enlarge the acceptance of the detector improving the reconstruction of 
decay modes with neutrinos ($B \to \tau \nu$, $B \to D^{(*)} \tau \nu$, $\tau$ decays, etc.).
The ability to separate the $B$ from the $D$ vertex would help rejecting the $q\bar{q}$ events 
($q = u, d, s, c$ quarks) and would open new analyses techniques for $B$ flavor tagging based on 
topological algorithms. This possibility has still to be quantitatively estimated.

\section{Tracking}
\label{sec:tracking}

Charged particle tracking is performed at \babar\ and \Belle through a
combination of silicon strip detectors and drift chamber. This design
allows a high resolution determination of the track parameters near
the interaction point while retaining good momentum resolution in the
gaseous drift chamber. Occupancies in the two systems are acceptable
with background currently experienced at the two $B$ Factories and we
expect that a similar detector design will be usable at \SuperB. 

A small radius vertexing device, directly mounted on the beam pipe,
will be required to improve vertex resolution. Monolithic active pixel
systems are being developed by several groups and are a very promising
technology for this application. In these devices the active detection
thickness is only of the order of 10-20\mum, and they can therefore be
thinned down to 50\mum, thus significantly recucing the amount of
material in the first detection layer.

The current silicon strip detector systems have four or five layers,
strip lengths between 5 and 20 cm, strip pitch between 50 and 200\mum,
and are readout through preamplifiers with shaping times between 100
and 400\ns. Given the time structure of events at \SuperB and the
projected low backgrounds, it should be possible to increase the
shaping time of the preamplifiers to 0.5-1\mus, thus improving the
signal to noise ratio and making it possible to reduce the detector
thickness from the current 300\mum to 200\mum. 

If the amount of material is reduced, decreasing the intrinsic
resolution from the current 12-20\mum to 5-10\mum through a finer
pitch readout would both benefit the momentum resolution, especially
for high momentum particles, and improve the angle measurement, which
is crucial for vertexing as well as for particle identification based
on \v Cerenkov angle determination.  The small angle region will require
special attention because of the high rate Bhabha
scattering. Depending on the specific choice of boost and angular
coverage a special section of silicon strips with finer segmentation
and/or shorter shaping time will be required. This section could be
organized in forward and backward disks or cones electrically
separated from the central part of the detector.

The \babar\ and \Belle drift chambers are conceptually similar but have
different dimensions: from 23.6 to 80.9\cm for \babar; from 16 to
114\cm for \Belle. They use He-based gas mixtures, with cell sizes from
12 to 18mm. Maximum drift time is around 500\ns and the resolution
about 100\mum\ in the best part of the cell. In preparation for
luminosity upgrades, \Belle has experimented with small cell (5.4\mm)
drift chamber for the inner layers of the CDC finding a significant
rate reduction but a deteriorated resolution (150\mum).

We expect that the reduction in cell size will not be needed at
\SuperB. On the contrary one could take advantage of the time
structure of the events and increase the drift time to up to 1\mus
with the intent of improving the resolution. A full optimization of
the drift chamber design will only be possible when detailed
background simulations are available.

\section{Particle Identification}
\label{sec:pid}

Particle identification is a central aspect of the criteria describing
the desired properties of a detector working at \SuperB.  Such a
detector system generally requires good tracking information to help
optimize the PID function, good start time information and generally
needs supplementary information, (such as $dE/dx$ from the energy loss
from the tracking chambers), to cover the full desired momentum range.

The \babar and \Belle detectors incorporate relatively complete PID
systems, which essentially provide the needed performance for their
respective environments. The \babar DIRC system has been particularly
robust, relatively insensitive to background, covers most of the
momentum range for all particle species, and has extremely good
misidentification properties. A new PID system based on the DIRC
technique but using modern photon detectors, such as a Fast Focusing DIRC or
a TOP, would provide more complete geometrical and momentum coverage,
and perhaps better rejection at the highest momenta.  The final choice
of collision energies at the SuperB Factory will define the highest
energy/momentum needed for pion/kaon separation.  It is also probable
that the End Cap regions will require more attention than has been
given for the current detectors.

The environment at the proposed SuperB Factory should not be a problem
for the PID systems described above. The beam currents are down a lot,
and the backgrounds should be much less severe than presently
experienced, although this will depend, in detail, on the design and
implementation of the final focus system The data acquisition problem
will be interesting, but quite manageable, requiring a pipelined
approach, but similar to the solution for other systems, and
achievable with commodity equipment.

More generally, the technologies of choice for PID in a new super B
factory detector appear to be two variations of the \v Cerenkov focusing
device, either in two or in three dimensions the DIRC of BaBar, a fast
DIRC, or the TOP of \Belle; and two systems that have special interest
for the end cap regions an aerogel proximity focusing device, and
possibly conventional time of flight with really good timing, (of
order 30 picoseconds) providing there is adequate momentum resolution
from the forward tracking devices.

Any new technology proposed to replace the established PID systems
will require a thorough and aggressive R\&D activity, proving not only
the device particle identification performance in real life
conditions, but also to demonstrate the photon detector lifetimes,
robustness and reliability, under realistic conditions.

The SLAC group is in the middle of beam test of the Fast Focusing Dirc
idea, with encouraging results.  We are also in the middle of testing
the micro-channel plate single photon detectors in magnetic fields up
to 15 KGauss, with encouraging results.

\section{Calorimetry}
\label{sec:calorimetry}

Although detailed background calculations for \SuperB have not yet
been done, the existing CsI(Tl) electromagnetic calorimeters of
\babar\ and \Belle should be suitable for use at the \SuperB MegaHertz
collision rate, at least for the main barrel sections. The long (1.3
$\mu$s) decay time of CsI(Tl) should not be a severe problem. The
radiation dose rate should be an order of magnitude smaller than at
the current $B$ factories. The barrel calorimeters of either current
detector should therefore be adequate without modification.

There may, however, be reason to consider upgrading the endcap
calorimeters. It is likely that the smaller size of beamline
elements inside the detector will allow extension of the solid angle
coverage for tracking and calorimetry. This would be beneficial to
all rare decay physics involving missing energy, including such
important topics as as measurement of $B\to\tau\nu$ or
$B\to K\nu\bar{\nu}$ branching fractions.

The cross section for Bhabha scattering events at 100 mrad in the
endcap region is more than an order of magnitude greater that it is
at 300 mrad; at small polar angles there will thus be a significant
number of events in which a Bhabha electron overlaps a hadronic
event. The most effective way to combat this is to use a photon
detection medium with faster response and shorter Moli\`{e}re
radius, to minimize the probability of temporal or spatial overlaps.
Development work on suitable crystals, cerium-doped lutetium
orthosilicate (LSO) and lutetium yttrium orthosilicate (LYSO), which
have a Mol\`{e}re radius 60\% of that of CsI, is underway at
Caltech. The two crystals are very similar, with LYSO somewhat
easier to grow and thus somewhat less expensive. LSO/LYSO is also
mechanically strong, and is not hygroscopic. A comparison of the
properties of CsI(Tl) and LSO/LYSO(Ce) is shown in
Table~\ref{tab:crystalcomp}.

\begin{table*}
  \caption{Comparison of the properties of CsI(Tl) and LSO/LYSO(Ce)}
  \label{tab:crystalcomp}
  \centering
   \begin{tabular}{lcc}
    \hline
          & CsI(Tl) & LSO/LYSO(Ce) \\ \hline
       Radiation length   (cm)   &  1.85  & 1.14 \\
      Moli\`{e}re radius (cm) & 3.5 & 2.3 \\
      Peak luminescence (nm) & 560 & 420 \\
      Decay time (ns) & 1300 & 42 \\
      Relative light yield & 1 & .65 \\
    \hline
  \end{tabular}
\end{table*}

A replacement of the existing endcaps in \Belle or replacement of the
existing forward endcap in \babar\ and the addition of a rear endcap
is thus likely to be a worthwhile upgrade. This would allow
extension of the solid angle coverage to the 100 mrad regime, and
minimize the number of events with overlapping Bhabhas. The \babar\
forward endcap calorimeter covers the solid angle down to 350 mrad;
there is no backward endcap. The forward calorimeter of \Belle covers
the region down to 200 mrad, while the backward endcap calorimeter
extends to 400 mrad.

\begin{figure*}[!htb]
\centering
\includegraphics[width=0.8\textwidth]{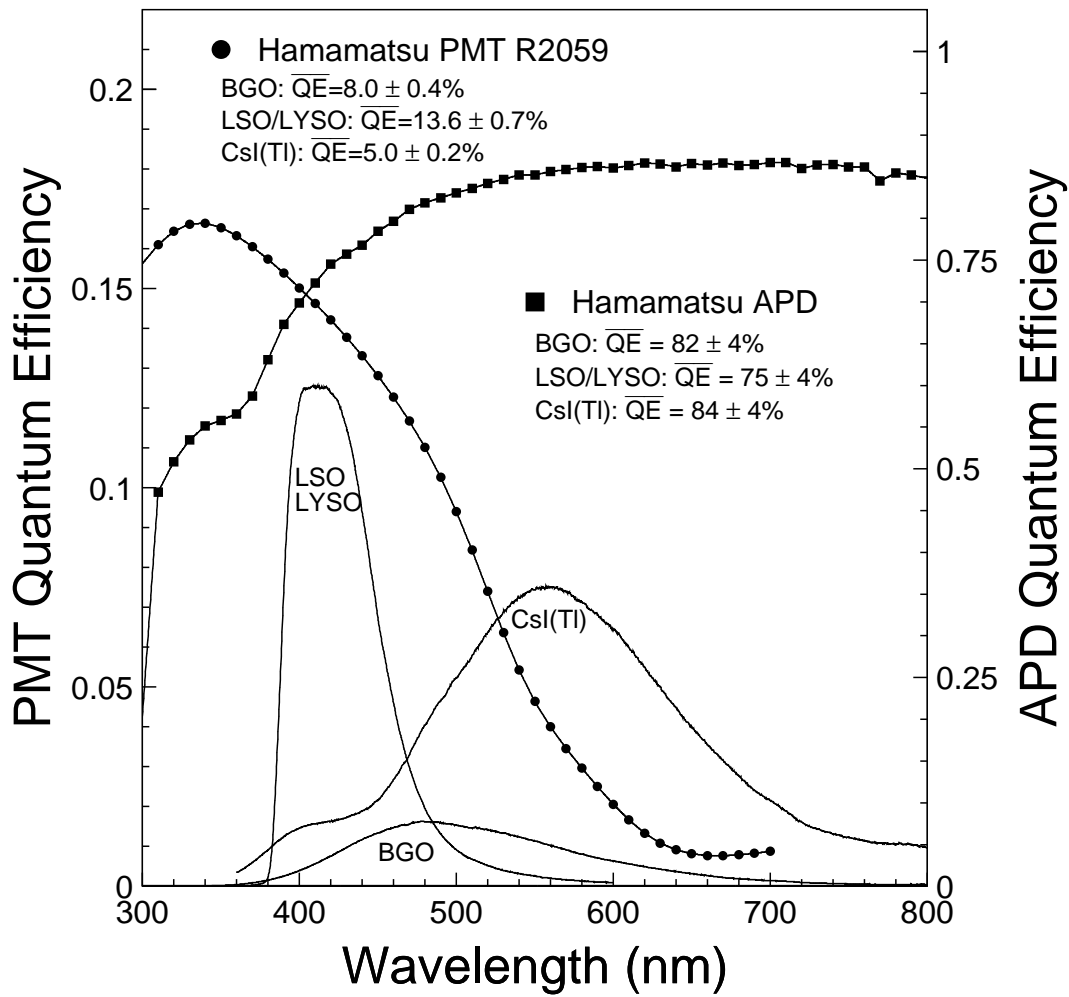} 
\caption{Emission
spectra of BGO, LSO/LYSO(Ce) and CsI(Tl) together with the quantum
efficiency of a Hamamatsu photomultiplier tube and avalanche
photodiode. The quantum efficiency averaged over the LSO/LYSO
spectrum is 75\% for the APD and 13.6\% for the PMT.}
\label{fig:spectrum}
\end{figure*}

The spectrum of scintillation light in LSO(Ce) peaks at 420 nm, has
a total light output 65\% of that of CsI(Tl), a very fast decay time
of 42 ns and is extremely radiation hard. The spectrum is
well-matched to solid state readout by an avalanche photodiode (APD)
or a conventional photodiode. A $^{137}$Cs spectrum using a
full-size LSO bar and a single Hamamatsu APD, as well as comparison
to a photomultiplier tube, is shown in Figure~\ref{fig:spectrum}.

A certain amount of additional R\&D is necessary to bring
LSO/LYSO(Ce) to a state in which one could build an actual
calorimeter. This R\&D involves optimization of the uniformity of
the cerium dopant, analysis and removal of trace impurities from the
basic salt, which are the source of a small amount of
phosphorescence observed under intense radiation doses, and further
optimization of the APD readout. There is already enough crystal
growth capacity to produce the endcap crystals in a few months.
Working with the crystal growers, we expect to be able to resolve
the remaining large crystal-related issues and to bring the price
into a more affordable range. These tasks could all be accomplished
within a year or two; an LSO(Ce) upgrade of the endcap regions of
either \babar\ or \Belle could thus be accomplished on the desired
time scale.

\section{Muon Detector}
\label{sec:muon}

The muon detection system at current $B$ Factories is realized by
instrumenting the flux return yoke of the magnets with large area
tracking detectors made of resistive plate chambers or limited
streamer tubes. These technologies have limited rate capabilities, but
with the current understanding of machine backgrounds it will be
possible to use these detectors at \SuperB. Special care must be taken
to shield the outermost layers of the muon detection systems from
radiation coming from the tunnel. This could be generated for instance
by synchrotron radiation backsplashing on far-away masking elements.
In addition, depending on the choices on the boost and on the minimum
angular coverage, the small angle portion of the detector will
probably require increased segmentation of the readout and/or separate
gas volumes. There seem to be no significant technological problem in
designing the muon detection system at \SuperB.

\section{Trigger-DAQ}
\label{sec:trigger}

The trigger and DAQ system for the detector at the SuperB Factory will
have the task of handling the large event and data rates
produced at a luminosity of $\lum = 10^{36}$.  The type of trigger
needed depends critically on the bunch collision frequency.  At
frequencies at or  below  $1$~MHz an online trigger would only
provide a factor of 10 or less in rejection, and at this level all
event rejection could be implemented instead in software triggers.  At
frequencies above $1$~MHz a hardware trigger would be desirable to
reduce the event read-out rate to the $100$~kHz level.  For any
configuration, highly pipelined readout in the DAQ system will be
necessary.  Any detector operating at  $\lum = 10^{36}$ will have 
DAQ rates of the order of $5$~GBytes/sec. This high
rate is, however, roughly an order of magnitude smaller than the front-end read-out
rates of the LHC experiments.  Extrapolations, including Moore's law
scaling for networks, disks, and CPUs, indicate that the trigger and
DAQ at SuperB will be comparable in difficulty to the current \babar
system.

%% file: accPart.tex
A Super B Factory, an asymmetric energy
\textit{e}\textit{$^{+}$}\textit{e}\textit{$^{-}$} collider with a
luminosity of order 10$^{36}$~cm$^{-2}$s$^{-1}$, can provide a
sensitive probe of new physics in the flavor sector of the Standard
Model. The success of the PEP-II and KEKB asymmetric colliders
\cite{bib:bib1,bib:bib2}
in producing unprecedented luminosity above 10$^{34}$~cm$^{-2}$s$^{-1
}$has taught us about the accelerator physics of asymmetric
\textit{e}\textit{$^{+}$}\textit{e}\textit{$^{-}$} colliders in a new
parameter regime. Furthermore, the success of the SLAC Linear Collider
\cite{bib:bib3} and the subsequent work on the International Linear
Collider \cite{bib:bib4}
allow a new Super-B collider to also incorporate linear collider
techniques. This note describes the initial parameters of a linearly
colliding asymmetric B-Factory collider at a luminosity of order
10$^{36}$~cm$^{-2}$s$^{-1}$.  Such a collider would produce an
integrated luminosity of about 10,000 fb$^{-1}$ (10 ab$^{-1}$) in a
running year (10$^{7}$~sec). Design studies are continuing to improve
these parameters.

\section{Design from past successes}

The construction and operation of modern multi-bunch
\textit{e}$^{+}$\textit{e}$^{-}$ colliders have brought about many
advances in accelerator physics in the area of high currents, complex
interaction regions, high beam-beam tune shifts, high power RF
systems, controlled beam instabilities, rapid injection rates, and
reliable uptimes ($\sim 95\%$).

The present successful B-Factories have proven that their design
concepts are valid:
\begin{enumerate}
\item Colliders with asymmetric energies can work. 
\item Beam-beam energy transparency conditions are weak.
\item Interaction regions with two energies can work. 
\item IR backgrounds can be handled successfully. 
\item High current RF systems can be operated (3 A$\times$1.8 A).  
\item Beam-beam parameters can reach 0.06 to 0.09. 
\item Injection rates are good and continuous injection is done in production. 
\item The electron cloud effect (ECI) can be managed. 
\item Bunch-by-bunch feedbacks at the 4 nsec spacing work well.
\end{enumerate}
Lessons learned from linear collider studies have also shown new successful concepts: 
\begin{enumerate}
\item Bunch energy and energy spread compensation of multiple bunches in a high power linac can be done. 
\item Small horizontal and vertical emittances can be produced in a damping ring with a short damping time. 
\item Superconducting linacs can be used with short high-charge bunches. 
\item Superconducting linacs can be used for beam energy recovery. 
\item Transverse beam kickers with fast switching times and excellent stability can be produced. 
\item Bunch length compression can be successfully performed.
\end{enumerate}
All of the above techniques will be incorporated in the design of a
future linear Super-B Factory collider.

\section{Design Status}

The concept of combining linear and circular collider ideas to make a
linear-circular B-Factory was discussed in the late 1980's, although
only circular B-Factories were built in the 1990's. Recent advances in
B-Factory performance and solid linear collider design progress has
reopened this design avenue.

The design presented here is very recent and on-going. There are new
ideas emerging on the weekly time scale and some time will be needed
to allow these new ideas to be incorporated into the ultimate
design. The parameters presented here are preliminary but with the
intent to be self-consistent.

\section{Luminosity}

The design of a 10$^{36}$~cm$^{-2}$s$^{-1}$
\textit{e}$^{+}$\textit{e}$^{-}$ collider combines extensions of the
design of the present \textit{B} Factories and linear collider
concepts to allow improved beam parameters to be achieved. The
luminosity L in an \textit{e}$^{+}$\textit{e}$^{-}$ collider is given
by the expression
\begin{eqnarray*}
L & = & \frac{N^+ N^- n_b f_c H_d}{4\pi\sigma_x\sigma_y} \\
\sigma & = & \sqrt{\beta\epsilon}
\end{eqnarray*}
where $n_{b}$ is the number of bunches, $f_{c}$ is
the frequency of collision of each bunch, $N$ is the number of particles
in the positron ($+$) and electron ($-$) bunches, $H_{d}$ is the
disruption enhancement factor from the collisions, $\sigma$ is the
beam size in the horizontal ($x$) and vertical ($y$) directions,
$\epsilon$ is the beam emittance and $\beta$ is the beta function (cm)
at the collision point for each plane.

\section{Collider concepts and layout}

\begin{figure}[!htb]
\includegraphics[angle=270,width=\linewidth]{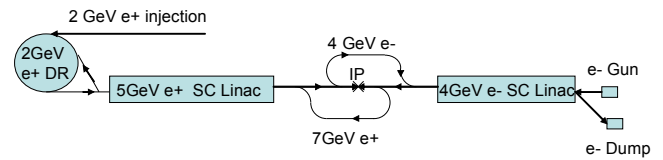}
\caption{Linearly colliding Super-B Factory layout}%
\label{fig:fig1a}
\end{figure}

\begin{figure}[!htb]
\includegraphics[angle=270,width=\linewidth]{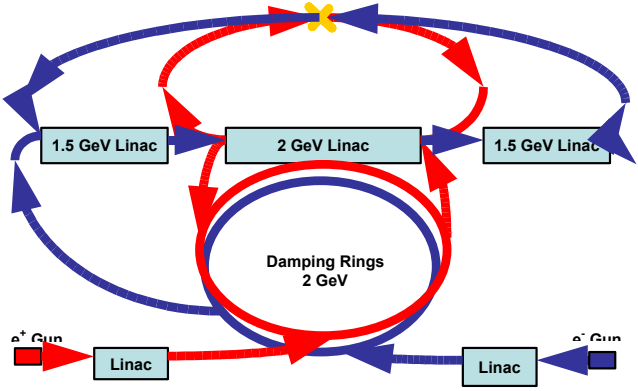}
\caption{Linearly colliding Super-B Factory layout}
\label{fig:fig1b}
\end{figure}

Schematic drawings of a Linear Super-B Factory are shown in
Figure~\ref{fig:fig1a} and \ref{fig:fig1b}.
The operation is described here. A positron bunch from a 2~GeV
damping ring is extracted and accelerated to 7~GeV in a
superconducting (SC) linac. Simultaneously, an electron bunch is
generated in a gun and accelerated in a separate SC linac to 4
GeV. The two bunches are prepared to collide in a transport line where
the bunch lengths are shortened. These bunches are focused to a small
spot at the collisions point and made to collide. The spent beams are
returned to their respective linacs with transport lines where they
return their energies to the SC accelerator. The 2~GeV positrons are
returned to the damping ring to restore the low emittances. The spent
electron beam is discarded. The process is repeated with the next
bunch. It is expected that each bunch will collide about 120 times
each second and that there will be about 10000 bunches. Thus, the
collision rate is about 1.2~MHz. A small electron linac and positron
source are used to replenish lost positrons in the colliding process
and natural beam lifetime. See Figure~\ref{fig:fig1a}.

An alternative (see Figure~\ref{fig:fig1b}) electron source is to use a 2~GeV
damping ring to store and collide electrons in a similar fashion to
positrons. This scheme would reduce the demands on the electron gun
but increase the site AC power.

Another alternative overall design could combine the two linacs into a
single unit, saving construction costs.

\section{Collision Parameters}

The parameters of the beam collisions are listed in
Table~\ref{tab:tab1}. The asymmetric energies are required to allow
tracking particle vertices in the collisions.

\begin{table}[!htb]
  \caption{Preliminary Super-B Factory collision parameters.}
  \label{tab:tab1}
  \begin{tabularx}{\linewidth}{|X|l|l|}\hline
    \textbf{Parameter} & \textbf{LEB} & \textbf{HEB} \\ \hline
    Beam Energy (GeV) & 4 & 7 \\ \hline
    Number of bunches & 10000 & 10000 \\ \hline
    Collision freq/bunch (Hz) & 120 & 120 \\ \hline
    IP energy spread (MeV) & 5 & 7 \\ \hline
    Particles /bunch $\times 10^{10}$ & 10 & 10 \\ \hline
    Time between collisions (msec) & 8.3 & 8.3 \\ \hline
    by* (mm) & 0.5 & 0.5 \\ \hline
    bx* (mm) & 22 & 22 \\ \hline
    Emittance (x/y) (nm) & 0.7/0.0016 & 0.7/0.0016 \\ \hline
    sz (mm) & 0.35 & 0.35 \\ \hline
    Lumi enchancement Hd & 1.07 & 1.07 \\ \hline
    Crossing angle(mrad) & 0 & 0 \\ \hline
    IP Horiz. size (mm) & 4 & 4 \\ \hline
    IP Vert. size (mm) & 0.028 & 0.028 \\ \hline
    Horizontal disruption & 1.7 & 0.9 \\ \hline
    Vertical disruption & 244 & 127 \\ \hline
    Luminosity (x1034/cm2/s) & 100 & 100 \\ \hline
  \end{tabularx}
\end{table}

\section{Beam-Beam Calculations}

The beam-beam interaction in a linear collider is basically the same
Coulomb interaction as in a storage ring collider, with extremely high
charge densities at IP, leading to very intense fields; since in this
case quantum behavior becomes important it is necessary to use a
beam-beam code to predict luminosities and related backgrounds. The
``classical'' effects of the beam-beam interaction are characterized
by a parameter called ``disruption'', which can be seen as the
equivalent to what the linear beam-beam tune shift is in storage
rings. Typical values for D in the vertical plane are less than 30 in
ILC, and more than 50 in a ``linearly colliding'' SuperB-Factory. The
horizontal Ds are kept near or below 1 to reduce energy spread in the
beam. The beam-beam interaction in such a regime can be highly non
linear and unstable, leading to loss of luminosity, rather than gain,
and to emittance blow-up. Since the beams must be recovered in this
scheme, emittance blow-up should be kept at minimum in order to
decrease the number of damping time necessary before the beams can
collide again.

Let's now recall some of the scaling laws that can help in the choice
of the collision parameters.

The beam-beam disruption is defined as:
\begin{eqnarray*}
  D_{x,y}^\pm = \frac{N^\mp\sigma_z^\mp}{%
    \gamma^\pm\sigma_{x,y}^\mp\left(\sigma_x^\mp+\sigma_y^\mp\right)}
\end{eqnarray*}
where $N$ is the number of particles in one bunch, $\sigma$$_{z}$ is the
bunch length, $\gamma$ is the beam energy in terms of electron mass,
$\sigma$$_{x}$ and $\sigma$$_{y}$ are the beam spot sizes at
collision. All the quantities refer to the opposite beam, except for
the beam energy factor.

On the other hand the luminosity is proportional to:
\begin{eqnarray*}
  L \propto \frac{N^2}{\left(\sigma_x\sigma_y\right)}
\end{eqnarray*}
and the center of mass (cm) energy spread during collision can be defined as:
\begin{eqnarray*}
  \sigma_E^{\text{cm}} \propto \frac{N^2}{\left(\sigma_x^2\sigma_z\right)}
  \propto \frac{D_x N}{\sigma_z^2}
  \propto \frac{L\sigma_y}{\left(\sigma_x\sigma_z\right)}
\end{eqnarray*}

For ``linearly colliding'' beams a large contribution to the energy
spread comes from the beam-beam interaction via the ``beamstrahlung'',
synchrotron radiation produced during collision. Due to the high
fields at the interaction the beams lose more energy and the cm energy
spread increases. This is an unwanted effect, since the \FourS is
relatively narrow, so the cm energy spread should be as small as
possible.

As it can be seen from the previous formulas there are conflicting
requirements for the collision parameters. In fact increasing the
number of particles gives higher luminosity but also higher energy
spread. Also, a short bunch gives less disruption and more luminosity,
since $\beta_{y}$$^{*}$ can be decreased without having hourglass
effect, but produces larger cm energy spread.

The strong-strong collision regime requires a simulation, since
analytical treatment is limited. Preliminary beam-beam studies have
been performed with the ``GuineaPig'' computer code by D. Schulte
(CERN) \cite{bib:bib5}, which includes backgrounds calculations, pinch effect,
kink instability, quantum effects, energy loss, and luminosity
spectrum.  This code has been intensely used for ILC studies of
beam-beam performances and backgrounds.

Some time has been spent in optimizing the ``simulation'' parameters,
such as the number of longitudinal slices, macro-particles, grid
sizes, etc... versus the computing time. Then an intensive study of
the luminosity as a function of number $N$, bunch length, beam spot
sizes, beam emittances and energy asymmetry has been performed, while
trying to keep small the cm energy spread and the outgoing beam
emittances.

Some preliminary conclusions can be drawn from the large number of
runs performed with different collision parameters:
\begin{itemize}

\item
the bunch length should be as short as possible, this allows to
increase $\beta_{y}$$^{*}$ and luminosity, and gives less disruption;

\item
given the maximum storable beam current in the Damping Ring the number
of bunches should be as small as possible, i.e.\ the number of
particles/bunch should be as high as possible (see for example
Fig. 2), compatibly with the increase of the cm energy spread;

\item
the horizontal emittance should be increased so to minimize D: in this
case less time is needed to damp the spent beams. The corresponding
luminosity loss can be recovered by increasing the collision
frequency;

\item
increasing the beam aspect ratio, i.e.\ having very flat beams, helps
to overcome the kink instability. As a result the spent beam
emittances are less disrupted, D$_{y}$ is smaller and the cm energy
spread is not affected by the interaction.
\end{itemize}

\begin{figure}[!htb]
  \includegraphics[angle=270,width=\linewidth]{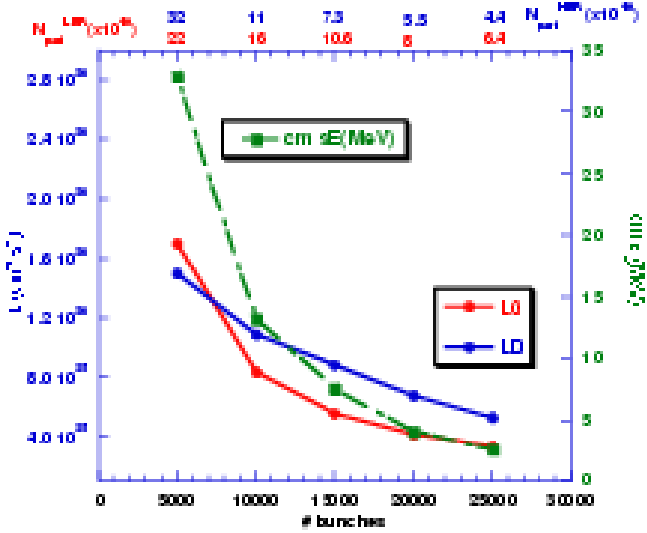}
  \caption{Luminosity and cm energy spread vs number of bunches for
    fixed current in the DR. In red is the geometric luminosity, in blue
    the disrupted one, in green the cm energy spread in MeV.}%
  \label{fig:fig2}
\end{figure}

\begin{figure}[!htb]
  \includegraphics[width=\linewidth]{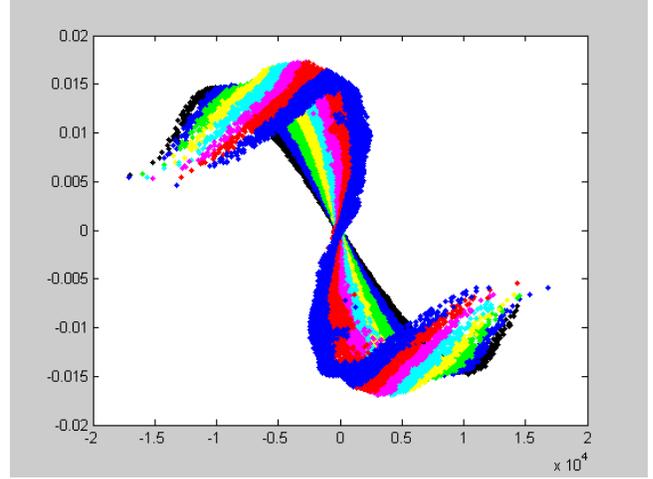}%
  \caption{Plot of the LER ($x,x'$) phase space. Each color
    refers to one longitudinal bunch slice.}
  \label{fig:fig3a}
\end{figure}

\begin{figure}[!htb]
  \includegraphics[width=\linewidth]{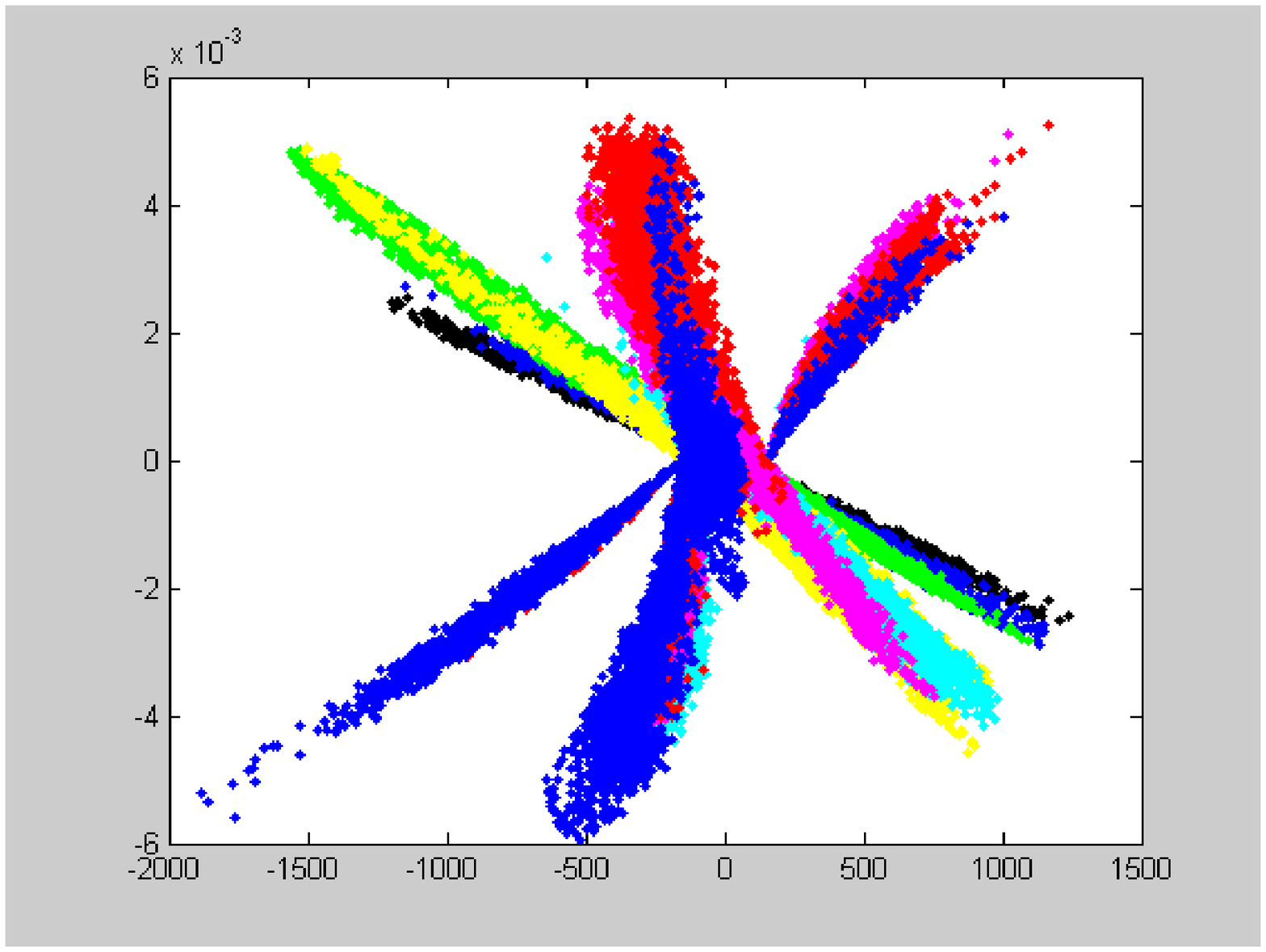}
  \caption{Plot of the LER ($y,y'$) phase space. Each color refers to one
    longitudinal bunch slice.}
  \label{fig:fig3b}
\end{figure}

As an example of spent beams emittances, in Figs.\ref{fig:fig3a} and
\ref{fig:fig3b} the ($x,x'$) and ($y,y'$) space phase plots after collision
for both beams, with the parameters in Table~\ref{tab:tab1}, are
shown. The different colors refer to different longitudinal bunch
slices, from the bunch head to the bunch tail.

\begin{figure}[!htb]
  \includegraphics[width=\linewidth]{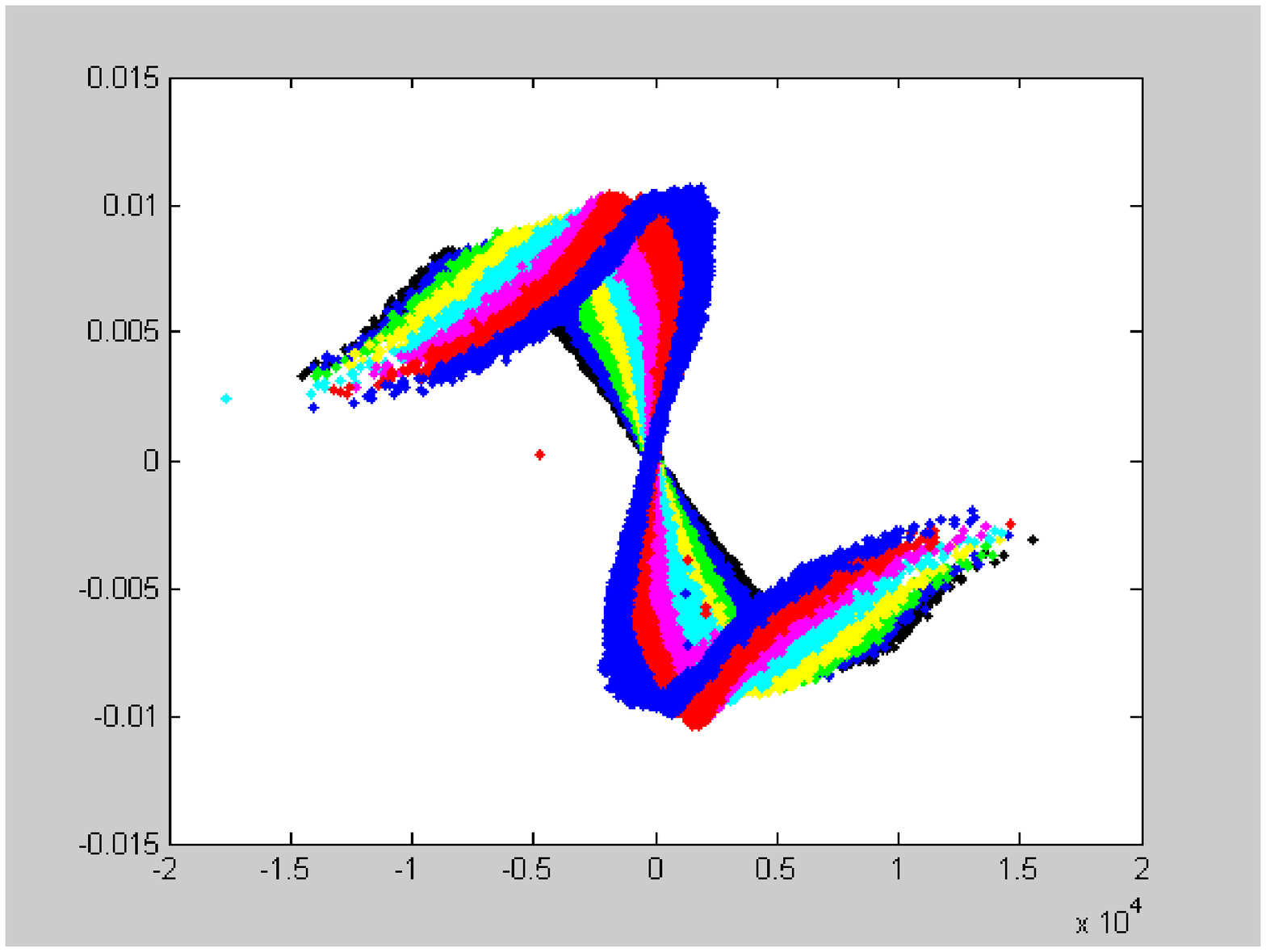}
  \caption{Plot of the HER ($x,x'$) phase space. Each color refers to one
    longitudinal bunch slice.}
  \label{fig:fig4a}
\end{figure}

\begin{figure}[!htb]
  \includegraphics[width=\linewidth]{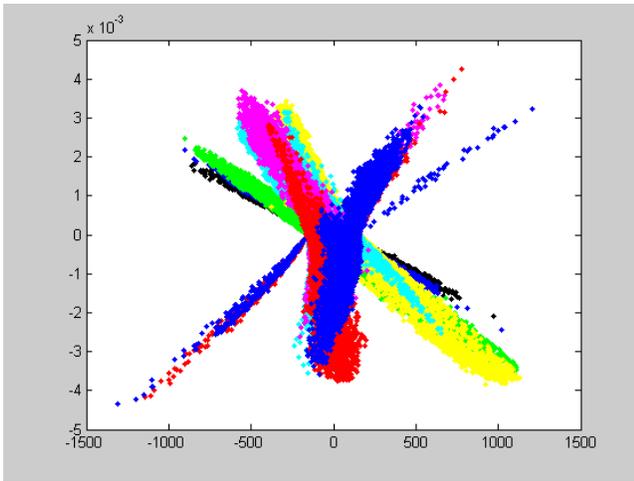}
  \caption{Plot of the HER ($y,y'$) phase space. Each color refers to one
    longitudinal bunch slice.}
  \label{fig:fig4b}
\end{figure}

\begin{figure}[!htb]
  \includegraphics[width=\linewidth]{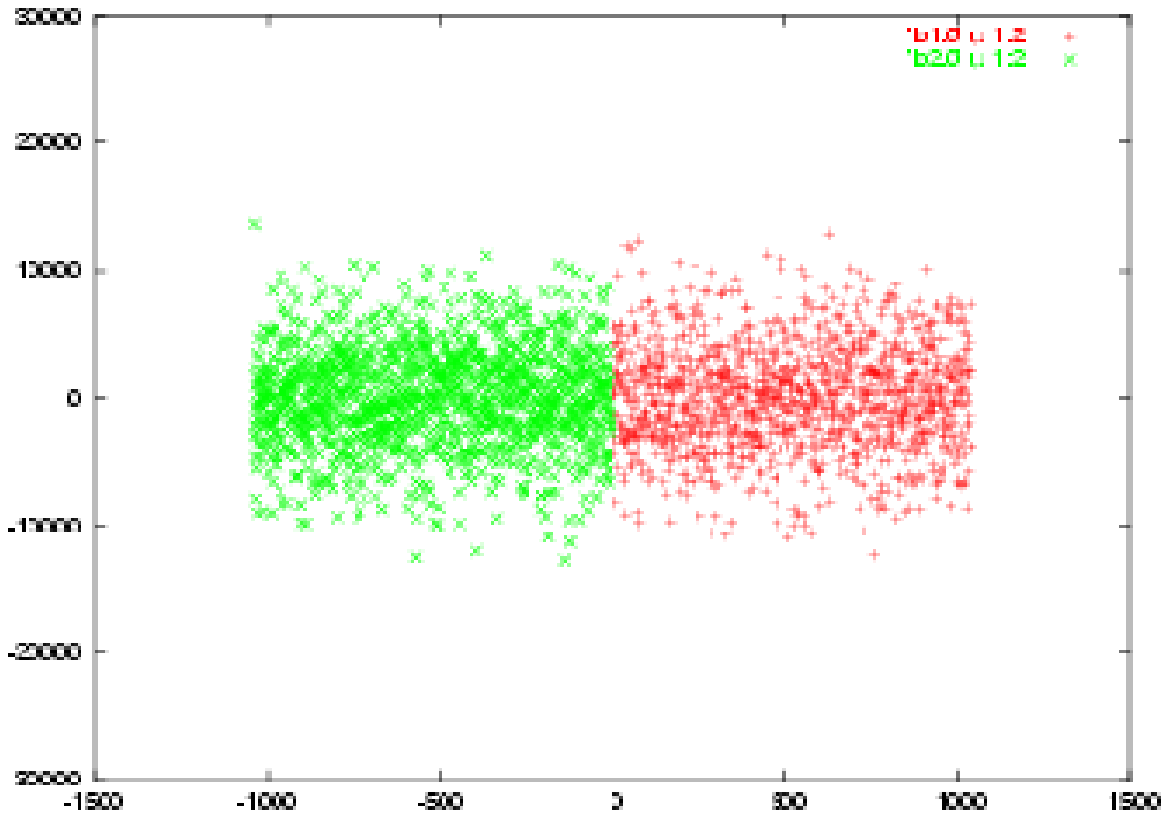} \\
  \includegraphics[width=\linewidth]{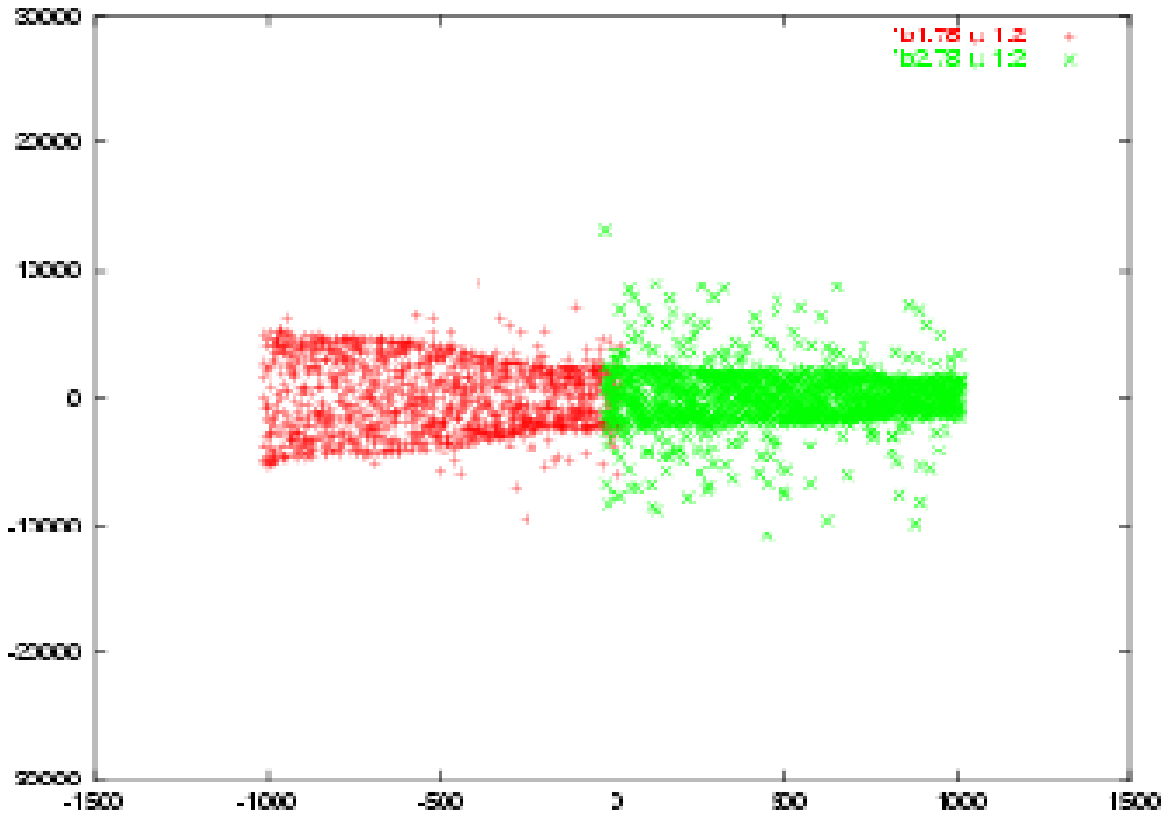}
  \caption{Beams before (top) and after (bottom) collision: horizontal
    vs longitudinal (z) distribution.
    LER particles are red, HER particles are green.}
  \label{fig:fig5}
\end{figure}

\begin{figure}[!htb]
  \includegraphics[width=\linewidth]{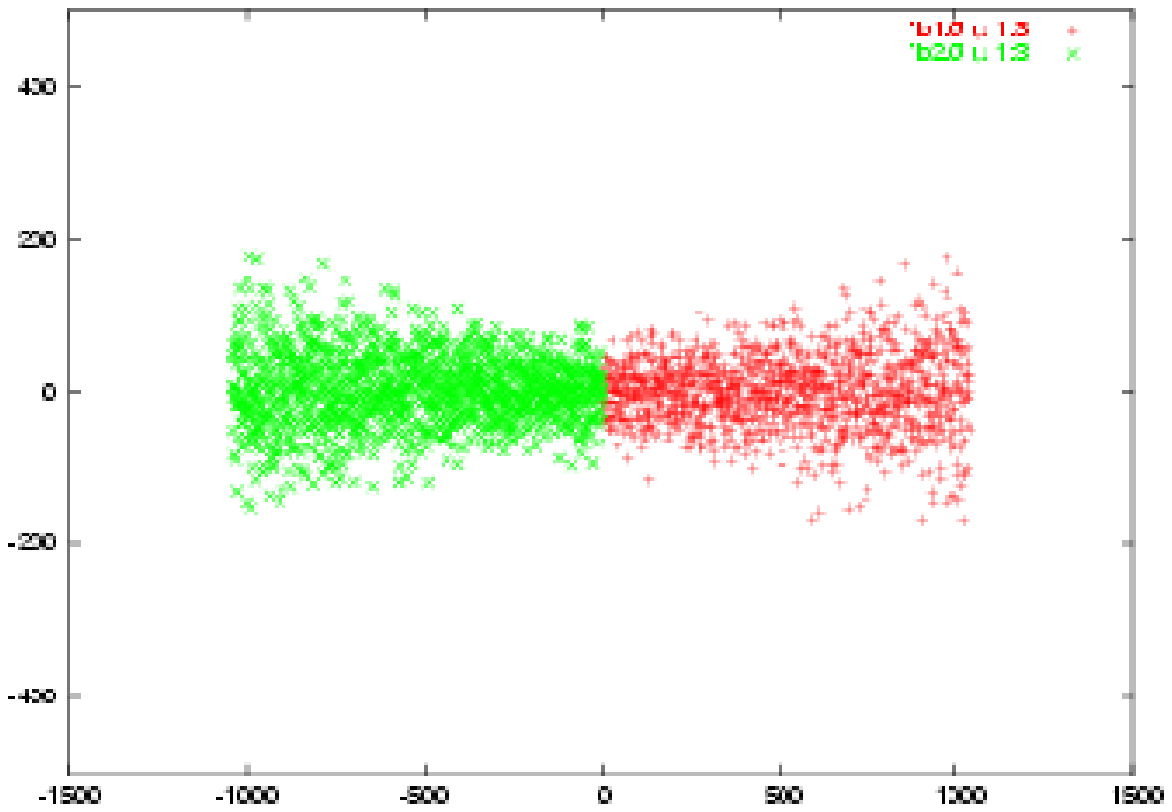} \\
  \includegraphics[width=\linewidth]{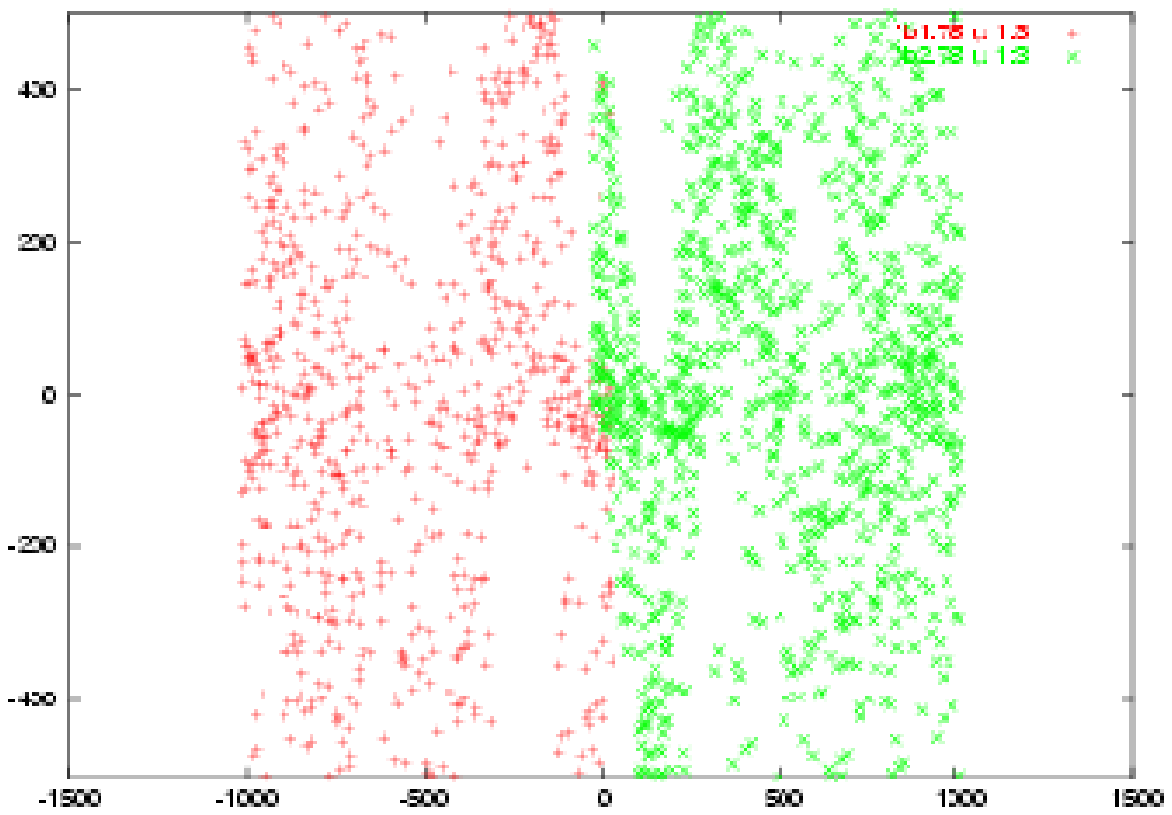}
  \caption{Beams before (top) and after (bottom) collision: vertical vs
    longitudinal (z) distribution.
    LER particles are red, HER particles are green.}
  \label{fig:fig6}
\end{figure}

In Figs.~\ref{fig:fig5} and \ref{fig:fig6} the simulated beams, before
and after collision, are shown. The low energy beam is red, the high
energy beam is green.

In this particular case the disruption is very high in the vertical
plane. This is what gives a large emittance blow-up. Also under study
are other possibilities to reduce the emittance growth due to the
collision.

The traveling focus scheme relaxes the requirements on the incoming
vertical emittance together with a reduced disruption during the
collision.

The four-beam DCI-like \cite{bib:bib6} beam charge compensation scheme (allowing
the beams to collide again before being sent back into the Linac), it
is also promising, since it could greatly reduce the disruption,
allowing much smaller IP sizes, together with very little emittance
growth, relaxing the requirements on beam current and damping time.

These studies are still in progress.

\section{Interaction Region Parameters}

The interaction region is being designed to leave the same
longitudinal free space as that presently used by \babar{} but
with superconducting quadrupole doublets as close to the interaction
region as possible.

Recent work at Brookhaven National Laboratory on precision conductor
placement of superconductors in large-bore low-field magnets has led
to quadrupoles in successful use in the interaction regions for the
HERA collider in Germany \cite{bib:bib7}.
A minor redesign of these magnets will
work well for the Super \textit{B} Factory.

The reduced energy asymmetry (7~GeV$\times$4~GeV) of the SuperB factory
makes it extremely difficult to separate the beams, especially if the
collision is head-on. However, the low emittance of the incoming beams
helps since the magnetic apertures can be small. On the other hand,
the exiting disrupted beams need large aperture magnets to contain the
beam with a minimum of loss. The reduced collision frequency (100~kHz
-- 1~MHz) of the SuperB Factory permits the introduction of pulsed
magnetic elements.

The plan to collide the two beams head-on and let the exiting beam
travel through the two final focus magnets of the other beam. The
exiting high-energy beam will be under-focused by the incoming
low-energy beam (LEB) final focus magnets. Two pulsed quadrupole
magnets will be used to add additional focusing for the outgoing
high-energy beam (HEB). These pulsed magnets will operate only on the
outgoing HEB and not be energized for the incoming LEB. The incoming
LEB will be steered to the final focus trajectory by a pulsed dipole
that will be outboard of the pulsed quadrupoles mentioned above. The
pulsed dipole on the incoming LEB allows the beams to be separated and
the out-going HEB then enters its own beam pipe.

On the other side of the IP the outgoing LEB is over-focused by the
incoming HEB final focus magnets. Again two pulsed quadrupole magnets
will be used to help capture the disrupted LEB. The LEB is then
steered out of the way by a pulsed dipole magnet allowing it to enter
its own beam pipe.

This design minimizes synchrotron radiation (SR) generated by the HEB
since the HEB runs straight through the Interaction Region (IR). The
SR fan from the incoming LEB pulsed dipole must be shielded from the
detector beam pipe. The remaining primary sources of SR that can be a
background for the detector, are generated by the final focus
quadrupoles for both incoming beams. This SR must be masked from the
detector beam pipe as well. The low emittance of the incoming beams
allows us to have a radius for the detector beam pipe of about 1~cm.

The masking for the detector beam pipe will be close to the beam pipe
in Z. This means that there will be scattered photons from the mask
tips and (most likely) photons striking the inside surfaces of the
masks from SR generated from the incoming beam on the other side of
the IP. This means that the detector Be beam pipe will need to have
the inside surface coated with a high-Z material (most likely gold) in
order to minimize the SR background in the detector.

Fig. 7 shows a possible IR layout. After passing through the collision
point (IP) HER and the LER beams are then steered away from the
incoming beam magnets with fast kicker dipole magnet. The RF quads
would turn on right after the incoming beam passes by.

\begin{figure}[!htb]
  \includegraphics[width=\linewidth]{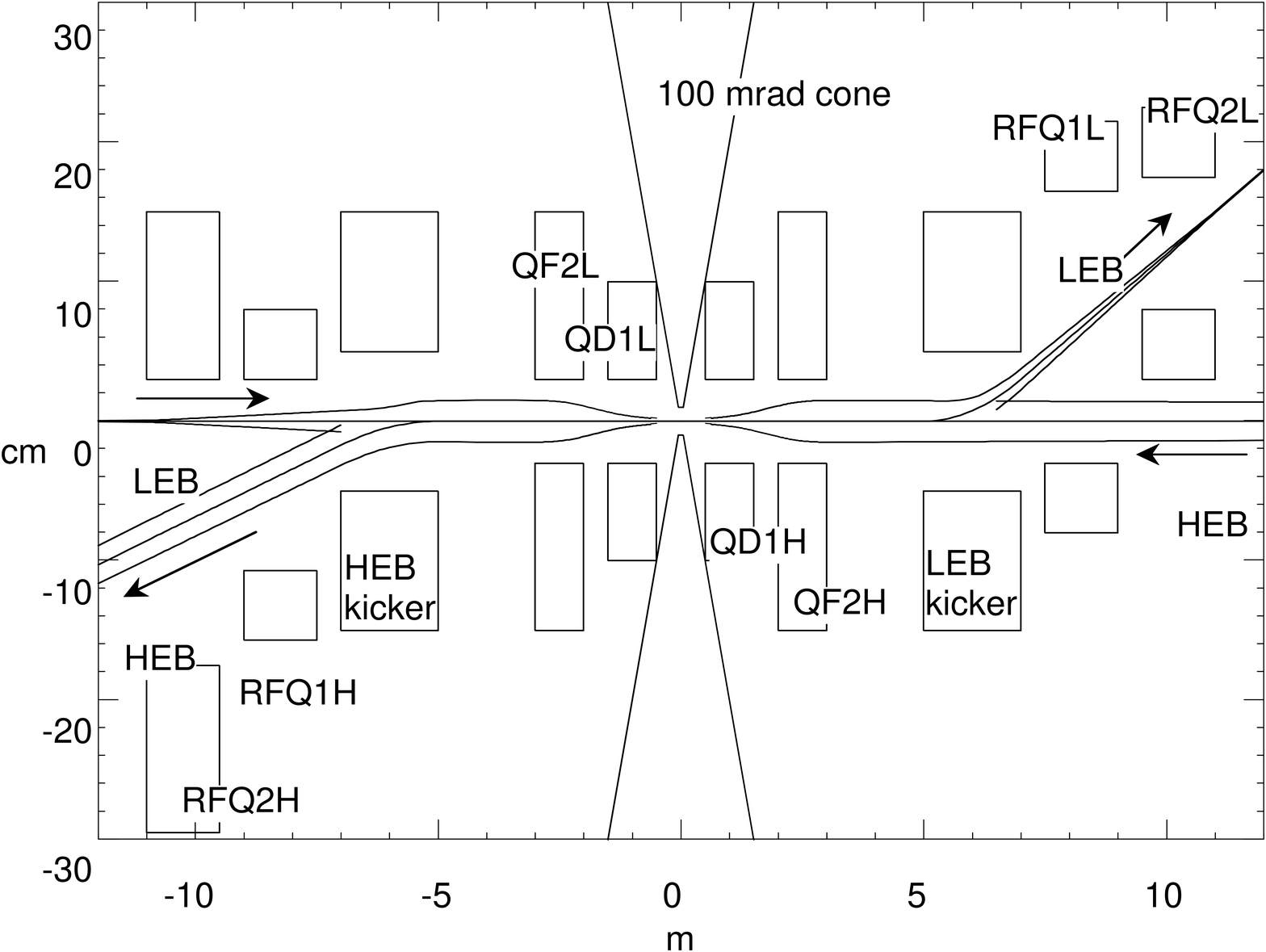}
  \caption{Plan view of a possible IR design.}
  \label{fig:fig7}
\end{figure}

\section{Linac Parameters}

The SC linear accelerator will have a design similar to the
International Linear Collider (ILC) SC structures with a frequency of
1428~MHz. Higher Order mode (HOM) damping will be needed similar to
the ILC. Since these structures will be powered at all times, an
accelerating gradient of about 8~MV/m is assumed here with a filling
factor of 0.7 similar to the ILC. Thus, a tunnel length of about 200~m
is needed for each GeV of acceleration including beam loading. The
klystron power needed here will be much lower than the ILC as the
energy is recovered for each bunch.

\section{Damping ring parameters}

The damping rings for each beam will have a total circumference if
6~km but two 3~km rings per particle type may be better. The first ring
will have a very short damping time to reduce the transverse
emittance. It will have a relatively large energy loss and large
equilibrium emittances. The second ring will have softer bends to have
smaller final emittances. Preliminary damping ring parameters are
listed in Table~\ref{tab:tab2}.\textit{}

Their characteristics are very similar to the ones studied for ILC
\cite{bib:bib4}.
The ILC-DR has the same length and emittances very similar to the
ones required for the Super-B. The ILC damping time is longer, so more
wigglers are needed here. A careful re-optimization of the ring
lattice has to be done in order to have a good dynamic aperture. The
additional wigglers together with the increased beam current
circulating in the ring (about a factor 5) will be a serious concern
for the electron-cloud instability. The other collective effects will
also be worse, despise the small benefit from the shorter damping
time. In addition we would like to have the rings operate at low
energy, to same power and cost, the design energy will be chosen after
a careful study, and probably will not be much lower than 2~GeV.

\begin{table}[!htb]
  \caption{Preliminary Damping Ring Parameters.}
  \label{tab:tab2}
  \begin{tabularx}{\linewidth}{|X|l|}\hline
    \textbf{Parameter} & \textbf{LEB} \\ \hline
    Energy  (GeV) & 2 \\ \hline
    Circumference (m) & 3000 \\ \hline
    Number of rings & 2 \\ \hline
    Average sync loss per turn MeV & 4.7 \\ \hline
    Total Synchrotron Rad Power MW/ring & 19 \\ \hline
    RF frequency (MHz) & 476 \\ \hline
    Vertical tune & 72.21 \\ \hline
    Horizontal tune & 76.29 \\ \hline
    Current (A) & 8 \\ \hline
    Bunches/ring & 5000 \\ \hline
    Particles per bunch & 1$\times$10$^{1}$$^{1}$ \\ \hline
    Ion gap (\%) & 1 \\ \hline
    Energy spread (\%) & 0.02 \\ \hline
    HER RF volts (MV) & 25 \\ \hline
    Longitudinal Damping time (msec) & 4.3 \\ \hline
    Emittance (x/y) (nm) & 0.7/0.0016 \\ \hline
    $\sigma$$_{z}$ (mm) & 3 \\ \hline
  \end{tabularx}
\end{table}

\section{Injector concept and parameters}

The injector for the SuperB will make up for lost particles during the
storage time in the damping rings and the losses from collisions. The
injector will be similar to the SLAC injector delivering about
5$\times$10$^{10}$ electrons or positrons per pulse at about 40~Hz each.

\section{Bunch compression}

The bunch compression system needs to compress the bunches between a
factor of 5 to 10 from 2-3~mm to 0.4-0.3~mm. Compression is done by
adding an head-tail energy correlation in each bunch and then passing
it through a transport line with dispersion. Since the initial energy
spread in the damping rings is very small, the induced energy spread
for compression will be about 1$\times$10$^{-3}$. Perhaps a multi-stage
compressor like in ILC will be required. The compressor could be
integrated with the SC accelerator in order to fully optimize the
bunch length, its profile (a rectangular distribution is probably
desired) and the final energy spread. The average dispersion needs to
be about 1~m over a bend angle of about 3~radians.

\section{Power Requirements}

The power required by a collider is the sum of a site base and the
accelerator operation. The damping ring power to replace the
synchrotron radiation loss will be the dominate factor in this Super-B
Factory.  The SC linac with energy recovery of the beams will not be a
large source.

\section{Synergy with ILC}

There are many similarities between this linear Super-B collider and
the ILC. The project described here will capitalize on R\&D projects
that have been concluded or are on-going with the ILC collaboration.

The damping rings between the two projects are very similar. Many of
the parameters are close such as energy (2-5~GeV), circumference
(3-6~km), bunch spacing (3-8 nsec), damping times (4-10~msec), and
emittances (3-10~nm horizontally and 0.002-0.001~nm vertically). Most
of the beam dynamics are due to multi-bunch effects which affect both
designs. The electron cloud effects will affect both rings in a
similar fashion. Both RF frequencies are in the range of 400 to 700~MHz.

The SC linacs have very similar characteristics including gradients
(5-25~MV/m), bunch spacing ($\sim 300$ nsec), and bunch charges
(1-10$\times$10$^{10}$).

The interaction regions have very similar characteristics with flat
beams and geometries. The ratio of IP beta functions are nearly the
same (10-30~mm horizontally and 0.3-1~mm vertically). The collimation
schemes should be similar. The possibility and techniques to use
traveling focusing will be similar, if needed. The chromatic
corrections of the final doublets using sextupoles will be the same.

All the beams will need bunch-by-bunch feedbacks to keep the beam
instabilities and beam-beam collisions under control. With the bunch
spacing very similar, the feedback kickers, digital controls, and beam
impedance remediation will have common designs. The IP feedback from
bunch-to-bunch will work exactly in the same way except that the
required SuperB Factory magnets will be much weaker. There will be
many opportunities to use feed-forward to correct bunch steering in
advance of the bunch arrival in the SuperB design.

\section{Other Upgrade possibilities}

Additional improvements are being considered for this design. 

1) A traveling focus scheme in the interaction region could help the beam-bean interaction and increase the luminosity or reduce the beam-beam blowup allowing the bunches to collide more frequently. 

2) A monochromator scheme could be used to reduce the energy spread tolerances in the interaction region and allow a simpler damping ring or enhanced luminosity. 

3) The SC linac could be ``doubled up'' to be used by more than one beam to reduce the construction cost. 

4) A four-beam collision scheme could be used to significantly reduce the effects of the beam-beam interaction allowing a much higher collision rate.

\section{Acknowledgments}

This document has come out of several recent SuperB workshops, with
the most recent one being at LNF (Frascati) on Nov.\ 11-12, 2005. We
appreciate very much discussions with the participants in these
workshops. We also appreciate discussions of parameters with members
of the ILC collaboration.